\documentclass[sigconf]{acmart}

\acmConference[ICSE 2024]{46th International Conference on Software Engineering}{April 2024}{Lisbon, Portugal}

\usepackage{xcolor}
\usepackage{xspace}
\usepackage{enumitem}
\usepackage{listings}

\newcommand{\DefMacro}[2]{\expandafter\newcommand\csname rmk-#1\endcsname{#2}}
\newcommand{\UseMacro}[1]{\csname rmk-#1\endcsname}
\newcommand{\InputWithSpace}[1]{\bgroup\def\arraystretch{1.02}\input{#1}\egroup}

\newcommand{\Tool}{OGO\xspace}
\newcommand{\ToolNeo}{OGO$^{Neo}$\xspace}
\newcommand{\ToolMemory}{OGO$^{Mem}$\xspace}
\newcommand{\OGO}{Object Graph Programming\xspace}

\newcommand{\ogo}{object graph programming\xspace}

\newcommand{\Cypher}{Cypher\xspace}
\newcommand{\openCypher}{openCypher\xspace}
\newcommand{\SQL}{SQL\xspace}
\newcommand{\LINQ}{LINQ\xspace}
\newcommand{\Neo}{Neo4j\xspace}
\newcommand{\Java}{Java\xspace}
\newcommand{\JavaVersion}{11.0.16\xspace}
\newcommand{\NeoVersion}{4.4.0\xspace}
\newcommand{\Antlr}{ANTLR4\xspace}

\newcommand{\JVMTI}{JVMTI\xspace}
\newcommand{\JNI}{JNI\xspace}
\newcommand{\JVM}{JVM\xspace}
\newcommand{\AST}{AST\xspace}
\newcommand{\Featherweight}{Featherweight\xspace}
\newcommand{\API}{API\xspace}

\newcommand{\FollowReferences}{FollowReferences\xspace}
\newcommand{\MyPara}[1]{\vspace{2pt}\noindent\textbf{#1}.}
\newcommand{\MyParaTwo}[1]{\vspace{0.5pt}\hspace{0.5pt}\textbf{#1}.}

\newcommand{\Project}{\textbf{Project}}

\newcommand{\SHA}{\textbf{SHA}}
\newcommand{\xNaive}{Naive\xspace}
\newcommand{\Naive}{\textbf{Naive}\xspace}
\newcommand{\xTraversePassedObjects}{FRO\xspace}
\newcommand{\TraversePassedObjects}{\textbf{+FRO}}
\newcommand{\xForceGC}{+WL+FGC\xspace}
\newcommand{\ForceGC}{\textbf{+FGC}}
\newcommand{\xAll}{+WL+FRO+FGC\xspace}
\newcommand{\All}{\textbf{+FRO+FGC}}

\newcommand{\xWhitelist}{WL}
\newcommand{\Whitelist}{\textbf{WL}}
\newcommand{\AntlrNaive}{\textbf{Naive}}

\newcommand{\AntlrForceGC}{\textbf{+FGC}}

\newcommand{\xAvg}{\textbf{Avg.}\xspace}
\newcommand{\xTotal}{\pmb{$\sum$}\xspace}

\newcommand{\Code}[1]{{\ifmmode{\mathtt{#1}}\else$\mathtt{#1}$\fi}}
\newcommand{\CodeIn}[1]{\texttt{\small #1}}
\newcommand{\Italicise}[1]{\textit{#1}}

\newcommand{\boundedq}{bounded query\xspace}

\newcommand{\unboundedq}{unbounded query\xspace}
\newcommand{\unboundedqs}{unbounded queries\xspace}

\newcommand{\assertjDB}{Assertj-db}
\newcommand{\commonsCli}{Cli}
\newcommand{\commonsCsv}{Csv}
\newcommand{\commonsLang}{Lang}
\newcommand{\commonsGeometry}{Geometry}
\newcommand{\jfreeChart}{Jfreechart}
\newcommand{\jsonJava}{Json-java}
\newcommand{\guava}{Guava\xspace}
\newcommand{\tcases}{Tcases}
\newcommand{\zipFourj}{Zip4j}
\newcommand{\jung}{JUNG\xspace}
\newcommand{\jcf}{JCF\xspace}

\newcommand{\guavaUnitTests}{611\xspace}
\newcommand{\jungUnitTests}{63\xspace}
\newcommand{\jcfUnitTests}{128\xspace}

\newcommand{\assertjDBSHA}{8aefa0f}
\newcommand{\commonsCliSHA}{0d06c4b}
\newcommand{\commonsCsvSHA}{1c551d9}
\newcommand{\commonsLangSHA}{dded8fd}
\newcommand{\commonsGeometrySHA}{6c6d046}
\newcommand{\jfreeChartSHA}{7cdbfbe}
\newcommand{\jsonJavaSHA}{31110b5}
\newcommand{\guavaSHA}{0ca124d}
\newcommand{\tcasesSHA}{2fccf74}
\newcommand{\zipFourjSHA}{fc3a258}

\newcommand{\assertjDBURL}{https://github.com/assertj/assertj-db}
\newcommand{\commonsCliURL}{https://github.com/apache/commons-cli}
\newcommand{\commonsCsvURL}{https://github.com/apache/commons-csv}
\newcommand{\commonsLangURL}{https://github.com/apache/commons-lang}
\newcommand{\commonsGeometryURL}{https://github.com/apache/commons-geometry}
\newcommand{\jfreeChartURL}{https://github.com/jfree/jfreechart}
\newcommand{\jsonJavaURL}{https://github.com/stleary/JSON-java}
\newcommand{\guavaURL}{https://github.com/google/guava}
\newcommand{\tcasesURL}{https://github.com/Cornutum/tcases}
\newcommand{\zipFourjURL}{https://github.com/srikanth-lingala/zip4j}

\newcommand{\assertjDBLOC}{78555}
\newcommand{\commonsCliLOC}{9256}
\newcommand{\commonsCsvLOC}{9910}
\newcommand{\commonsLangLOC}{86960}
\newcommand{\commonsGeometryLOC}{79996}
\newcommand{\jfreeChartLOC}{134399}
\newcommand{\jsonJavaLOC}{14829}
\newcommand{\guavaLOC}{367861}
\newcommand{\tcasesLOC}{482581}
\newcommand{\zipFourjLOC}{15525}

\newcommand{\assertjDBNumAssertions}{9}
\newcommand{\commonsCliNumAssertions}{27}
\newcommand{\commonsCsvNumAssertions}{12}
\newcommand{\commonsLangNumAssertions}{23}
\newcommand{\commonsGeometryNumAssertions}{95}
\newcommand{\jfreeChartNumAssertions}{10}
\newcommand{\jsonJavaNumAssertions}{10}
\newcommand{\guavaNumAssertions}{5}
\newcommand{\tcasesNumAssertions}{26}
\newcommand{\zipFourjNumAssertions}{18}

\newcommand{\ReductionAntlrFGCNeoAll}{\textbf{Redn.}\xspace}

\newcommand{\NaiveNeoTotal}{2698057}
\newcommand{\WlNeoTotal}{64197}

\newcommand{\FRONeoTotal}{64200}
\newcommand{\FGCNeoTotal}{22829}
\newcommand{\AllNeoTotal}{22197}
\newcommand{\NaiveAntlrTotal}{3579}
\newcommand{\WlAntlrTotal}{3592}
\newcommand{\FGCAntlrTotal}{1642}

\newcommand{\NaiveNeoAverage}{269806}
\newcommand{\WlNeoAverage}{6419}

\newcommand{\FRONeoAverage}{6420}
\newcommand{\FGCNeoAverage}{2283}
\newcommand{\AllNeoAverage}{2219}
\newcommand{\NaiveAntlrAverage}{358}
\newcommand{\WlAntlrAverage}{359}
\newcommand{\FGCAntlrAverage}{164}

\newcommand{\urlsymbol}{\includegraphics[width=.8em]{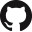}}

\newcommand{\TableFontSize}{\footnotesize}

\newcommand{\PerformanceCaption}{End-to-end execution time (ms) of assertions re-written using \Tool with different optimizations for external database and in-memory query executions.\vspace{-3pt}}

\DefMacro{assertjDBTraversePassedObjectsAvg}{7092}
\DefMacro{assertjDBTraversePassedObjectsNumRuns}{50}
\DefMacro{assertjDBTraversePassedObjectsStdev}{862}
\DefMacro{assertjDBTraversePassedObjectsAssertQuery}{227}
\DefMacro{assertjDBAllAvg}{4861}
\DefMacro{assertjDBAllNumRuns}{50}
\DefMacro{assertjDBAllStdev}{995}
\DefMacro{assertjDBAllAssertQuery}{223}
\DefMacro{assertjDBWhitelistAvg}{7063}
\DefMacro{assertjDBWhitelistNumRuns}{50}
\DefMacro{assertjDBWhitelistStdev}{505}
\DefMacro{assertjDBWhitelistAssertQuery}{223}
\DefMacro{assertjDBInstanceInfoReallocAvg}{7033}
\DefMacro{assertjDBInstanceInfoReallocNumRuns}{50}
\DefMacro{assertjDBInstanceInfoReallocStdev}{925}
\DefMacro{assertjDBInstanceInfoReallocAssertQuery}{223}
\DefMacro{assertjDBNoneAvg}{832652}
\DefMacro{assertjDBNoneNumRuns}{2}
\DefMacro{assertjDBNoneAssertQuery}{112}
\DefMacro{assertjDBForceGCAvg}{4799}
\DefMacro{assertjDBForceGCNumRuns}{50}
\DefMacro{assertjDBForceGCStdev}{659}
\DefMacro{assertjDBForceGCAssertQuery}{223}
\DefMacro{assertjDBCacheInstanceInfoAvg}{7156}
\DefMacro{assertjDBCacheInstanceInfoNumRuns}{50}
\DefMacro{assertjDBCacheInstanceInfoStdev}{1028}
\DefMacro{assertjDBCacheInstanceInfoAssertQuery}{225}
\DefMacro{assertjDBAntlrWhitelistAvg}{860}
\DefMacro{assertjDBAntlrWhitelistNumRuns}{50}
\DefMacro{assertjDBAntlrWhitelistStdev}{30}
\DefMacro{assertjDBAntlrNoneAvg}{851}
\DefMacro{assertjDBAntlrNoneNumRuns}{50}
\DefMacro{assertjDBAntlrNoneStdev}{35}
\DefMacro{assertjDBAntlrForceGCAvg}{248}
\DefMacro{assertjDBAntlrForceGCNumRuns}{50}
\DefMacro{assertjDBAntlrForceGCStdev}{12}
\DefMacro{assertjDBAllNumObjects}{699059}
\DefMacro{assertjDBWhitelistNumObjects}{4902394}
\DefMacro{assertjDBForceGCNumObjects}{109390}
\DefMacro{assertjDBAntlrWhitelistNumObjects}{985460}
\DefMacro{assertjDBAntlrNoneNumObjects}{974226}
\DefMacro{assertjDBAntlrForceGCNumObjects}{109924}
\DefMacro{assertjDBAntlrNumObjectsReduction}{88.85}
\DefMacro{assertjDBAntlrFGCNeoAll}{95}

\DefMacro{commonsCliTraversePassedObjectsAvg}{1982}
\DefMacro{commonsCliTraversePassedObjectsNumRuns}{50}
\DefMacro{commonsCliTraversePassedObjectsStdev}{277}
\DefMacro{commonsCliTraversePassedObjectsAssertQuery}{141}
\DefMacro{commonsCliAllAvg}{1880}
\DefMacro{commonsCliAllNumRuns}{50}
\DefMacro{commonsCliAllStdev}{268}
\DefMacro{commonsCliAllAssertQuery}{142}
\DefMacro{commonsCliWhitelistAvg}{1987}
\DefMacro{commonsCliWhitelistNumRuns}{50}
\DefMacro{commonsCliWhitelistStdev}{243}
\DefMacro{commonsCliWhitelistAssertQuery}{139}
\DefMacro{commonsCliInstanceInfoReallocAvg}{1961}
\DefMacro{commonsCliInstanceInfoReallocNumRuns}{50}
\DefMacro{commonsCliInstanceInfoReallocStdev}{261}
\DefMacro{commonsCliInstanceInfoReallocAssertQuery}{141}
\DefMacro{commonsCliNoneAvg}{99798}
\DefMacro{commonsCliNoneNumRuns}{2}
\DefMacro{commonsCliNoneAssertQuery}{74}
\DefMacro{commonsCliForceGCAvg}{1900}
\DefMacro{commonsCliForceGCNumRuns}{50}
\DefMacro{commonsCliForceGCStdev}{272}
\DefMacro{commonsCliForceGCAssertQuery}{142}
\DefMacro{commonsCliCacheInstanceInfoAvg}{1968}
\DefMacro{commonsCliCacheInstanceInfoNumRuns}{50}
\DefMacro{commonsCliCacheInstanceInfoStdev}{195}
\DefMacro{commonsCliCacheInstanceInfoAssertQuery}{141}
\DefMacro{commonsCliAntlrWhitelistAvg}{226}
\DefMacro{commonsCliAntlrWhitelistNumRuns}{50}
\DefMacro{commonsCliAntlrWhitelistStdev}{24}
\DefMacro{commonsCliAntlrNoneAvg}{224}
\DefMacro{commonsCliAntlrNoneNumRuns}{50}
\DefMacro{commonsCliAntlrNoneStdev}{25}
\DefMacro{commonsCliAntlrForceGCAvg}{160}
\DefMacro{commonsCliAntlrForceGCNumRuns}{50}
\DefMacro{commonsCliAntlrForceGCStdev}{13}
\DefMacro{commonsCliAllNumObjects}{607584}
\DefMacro{commonsCliWhitelistNumObjects}{4065234}
\DefMacro{commonsCliForceGCNumObjects}{584752}
\DefMacro{commonsCliAntlrWhitelistNumObjects}{203791}
\DefMacro{commonsCliAntlrNoneNumObjects}{201989}
\DefMacro{commonsCliAntlrForceGCNumObjects}{67022}
\DefMacro{commonsCliAntlrNumObjectsReduction}{67.11}
\DefMacro{commonsCliAntlrFGCNeoAll}{91}

\DefMacro{commonsCsvTraversePassedObjectsAvg}{2713}
\DefMacro{commonsCsvTraversePassedObjectsNumRuns}{50}
\DefMacro{commonsCsvTraversePassedObjectsStdev}{249}
\DefMacro{commonsCsvTraversePassedObjectsAssertQuery}{142}
\DefMacro{commonsCsvAllAvg}{2171}
\DefMacro{commonsCsvAllNumRuns}{50}
\DefMacro{commonsCsvAllStdev}{424}
\DefMacro{commonsCsvAllAssertQuery}{145}
\DefMacro{commonsCsvWhitelistAvg}{2867}
\DefMacro{commonsCsvWhitelistNumRuns}{50}
\DefMacro{commonsCsvWhitelistStdev}{348}
\DefMacro{commonsCsvWhitelistAssertQuery}{145}
\DefMacro{commonsCsvInstanceInfoReallocAvg}{2771}
\DefMacro{commonsCsvInstanceInfoReallocNumRuns}{50}
\DefMacro{commonsCsvInstanceInfoReallocStdev}{406}
\DefMacro{commonsCsvInstanceInfoReallocAssertQuery}{146}
\DefMacro{commonsCsvNoneAvg}{448403}
\DefMacro{commonsCsvNoneNumRuns}{2}
\DefMacro{commonsCsvNoneAssertQuery}{74}
\DefMacro{commonsCsvForceGCAvg}{2291}
\DefMacro{commonsCsvForceGCNumRuns}{50}
\DefMacro{commonsCsvForceGCStdev}{460}
\DefMacro{commonsCsvForceGCAssertQuery}{147}
\DefMacro{commonsCsvCacheInstanceInfoAvg}{2847}
\DefMacro{commonsCsvCacheInstanceInfoNumRuns}{50}
\DefMacro{commonsCsvCacheInstanceInfoStdev}{473}
\DefMacro{commonsCsvCacheInstanceInfoAssertQuery}{145}
\DefMacro{commonsCsvAntlrWhitelistAvg}{318}
\DefMacro{commonsCsvAntlrWhitelistNumRuns}{50}
\DefMacro{commonsCsvAntlrWhitelistStdev}{17}
\DefMacro{commonsCsvAntlrNoneAvg}{319}
\DefMacro{commonsCsvAntlrNoneNumRuns}{50}
\DefMacro{commonsCsvAntlrNoneStdev}{20}
\DefMacro{commonsCsvAntlrForceGCAvg}{178}
\DefMacro{commonsCsvAntlrForceGCNumRuns}{50}
\DefMacro{commonsCsvAntlrForceGCStdev}{15}
\DefMacro{commonsCsvAllNumObjects}{66009}
\DefMacro{commonsCsvWhitelistNumObjects}{116430}
\DefMacro{commonsCsvForceGCNumObjects}{66018}
\DefMacro{commonsCsvAntlrWhitelistNumObjects}{243383}
\DefMacro{commonsCsvAntlrNoneNumObjects}{242494}
\DefMacro{commonsCsvAntlrForceGCNumObjects}{72131}
\DefMacro{commonsCsvAntlrNumObjectsReduction}{70.36}
\DefMacro{commonsCsvAntlrFGCNeoAll}{92}

\DefMacro{commonsLangTraversePassedObjectsAvg}{1724}
\DefMacro{commonsLangTraversePassedObjectsNumRuns}{50}
\DefMacro{commonsLangTraversePassedObjectsStdev}{675}
\DefMacro{commonsLangTraversePassedObjectsAssertQuery}{109}
\DefMacro{commonsLangAllAvg}{1437}
\DefMacro{commonsLangAllNumRuns}{50}
\DefMacro{commonsLangAllStdev}{148}
\DefMacro{commonsLangAllAssertQuery}{108}
\DefMacro{commonsLangWhitelistAvg}{1718}
\DefMacro{commonsLangWhitelistNumRuns}{50}
\DefMacro{commonsLangWhitelistStdev}{609}
\DefMacro{commonsLangWhitelistAssertQuery}{108}
\DefMacro{commonsLangInstanceInfoReallocAvg}{1683}
\DefMacro{commonsLangInstanceInfoReallocNumRuns}{50}
\DefMacro{commonsLangInstanceInfoReallocStdev}{583}
\DefMacro{commonsLangInstanceInfoReallocAssertQuery}{109}
\DefMacro{commonsLangNoneAvg}{255390}
\DefMacro{commonsLangNoneNumRuns}{2}
\DefMacro{commonsLangNoneAssertQuery}{218}
\DefMacro{commonsLangForceGCAvg}{1510}
\DefMacro{commonsLangForceGCNumRuns}{50}
\DefMacro{commonsLangForceGCStdev}{273}
\DefMacro{commonsLangForceGCAssertQuery}{108}
\DefMacro{commonsLangCacheInstanceInfoAvg}{1718}
\DefMacro{commonsLangCacheInstanceInfoNumRuns}{50}
\DefMacro{commonsLangCacheInstanceInfoStdev}{527}
\DefMacro{commonsLangCacheInstanceInfoAssertQuery}{108}
\DefMacro{commonsLangAntlrWhitelistAvg}{286}
\DefMacro{commonsLangAntlrWhitelistNumRuns}{50}
\DefMacro{commonsLangAntlrWhitelistStdev}{8}
\DefMacro{commonsLangAntlrNoneAvg}{287}
\DefMacro{commonsLangAntlrNoneNumRuns}{50}
\DefMacro{commonsLangAntlrNoneStdev}{8}
\DefMacro{commonsLangAntlrForceGCAvg}{148}
\DefMacro{commonsLangAntlrForceGCNumRuns}{50}
\DefMacro{commonsLangAntlrForceGCStdev}{10}
\DefMacro{commonsLangAllNumObjects}{717697}
\DefMacro{commonsLangWhitelistNumObjects}{4043503}
\DefMacro{commonsLangForceGCNumObjects}{641287}
\DefMacro{commonsLangAntlrWhitelistNumObjects}{317674}
\DefMacro{commonsLangAntlrNoneNumObjects}{317848}
\DefMacro{commonsLangAntlrForceGCNumObjects}{64045}
\DefMacro{commonsLangAntlrNumObjectsReduction}{79.84}
\DefMacro{commonsLangAntlrFGCNeoAll}{90}

\DefMacro{commonsGeometryTraversePassedObjectsAvg}{1397}
\DefMacro{commonsGeometryTraversePassedObjectsNumRuns}{50}
\DefMacro{commonsGeometryTraversePassedObjectsStdev}{71}
\DefMacro{commonsGeometryTraversePassedObjectsAssertQuery}{69}
\DefMacro{commonsGeometryAllAvg}{1267}
\DefMacro{commonsGeometryAllNumRuns}{50}
\DefMacro{commonsGeometryAllStdev}{73}
\DefMacro{commonsGeometryAllAssertQuery}{68}
\DefMacro{commonsGeometryWhitelistAvg}{1407}
\DefMacro{commonsGeometryWhitelistNumRuns}{50}
\DefMacro{commonsGeometryWhitelistStdev}{60}
\DefMacro{commonsGeometryWhitelistAssertQuery}{68}
\DefMacro{commonsGeometryInstanceInfoReallocAvg}{1370}
\DefMacro{commonsGeometryInstanceInfoReallocNumRuns}{50}
\DefMacro{commonsGeometryInstanceInfoReallocStdev}{71}
\DefMacro{commonsGeometryInstanceInfoReallocAssertQuery}{69}
\DefMacro{commonsGeometryForceGCAvg}{1321}
\DefMacro{commonsGeometryForceGCNumRuns}{50}
\DefMacro{commonsGeometryForceGCStdev}{83}
\DefMacro{commonsGeometryForceGCAssertQuery}{68}
\DefMacro{commonsGeometryCacheInstanceInfoAvg}{1396}
\DefMacro{commonsGeometryCacheInstanceInfoNumRuns}{50}
\DefMacro{commonsGeometryCacheInstanceInfoStdev}{82}
\DefMacro{commonsGeometryCacheInstanceInfoAssertQuery}{69}
\DefMacro{commonsGeometryNoneAvg}{OOM}
\DefMacro{commonsGeometryAntlrWhitelistAvg}{257}
\DefMacro{commonsGeometryAntlrWhitelistNumRuns}{50}
\DefMacro{commonsGeometryAntlrWhitelistStdev}{8}
\DefMacro{commonsGeometryAntlrNoneAvg}{258}
\DefMacro{commonsGeometryAntlrNoneNumRuns}{50}
\DefMacro{commonsGeometryAntlrNoneStdev}{8}
\DefMacro{commonsGeometryAntlrForceGCAvg}{145}
\DefMacro{commonsGeometryAntlrForceGCNumRuns}{50}
\DefMacro{commonsGeometryAntlrForceGCStdev}{4}
\DefMacro{commonsGeometryAllNumObjects}{273830}
\DefMacro{commonsGeometryWhitelistNumObjects}{273830}
\DefMacro{commonsGeometryForceGCNumObjects}{65491}
\DefMacro{commonsGeometryAntlrWhitelistNumObjects}{266485}
\DefMacro{commonsGeometryAntlrNoneNumObjects}{266690}
\DefMacro{commonsGeometryAntlrForceGCNumObjects}{63814}
\DefMacro{commonsGeometryAntlrNumObjectsReduction}{76.05}
\DefMacro{commonsGeometryAntlrFGCNeoAll}{89}

\DefMacro{guavaTraversePassedObjectsAvg}{2784}
\DefMacro{guavaTraversePassedObjectsNumRuns}{50}
\DefMacro{guavaTraversePassedObjectsStdev}{387}
\DefMacro{guavaTraversePassedObjectsAssertQuery}{80}
\DefMacro{guavaAllAvg}{2408}
\DefMacro{guavaAllNumRuns}{50}
\DefMacro{guavaAllStdev}{65}
\DefMacro{guavaAllAssertQuery}{75}
\DefMacro{guavaWhitelistAvg}{2804}
\DefMacro{guavaWhitelistNumRuns}{50}
\DefMacro{guavaWhitelistStdev}{686}
\DefMacro{guavaWhitelistAssertQuery}{78}
\DefMacro{guavaInstanceInfoReallocAvg}{2687}
\DefMacro{guavaInstanceInfoReallocNumRuns}{50}
\DefMacro{guavaInstanceInfoReallocStdev}{97}
\DefMacro{guavaInstanceInfoReallocAssertQuery}{78}
\DefMacro{guavaNoneAvg}{175658}
\DefMacro{guavaNoneNumRuns}{2}
\DefMacro{guavaNoneAssertQuery}{223}
\DefMacro{guavaForceGCAvg}{2458}
\DefMacro{guavaForceGCNumRuns}{50}
\DefMacro{guavaForceGCStdev}{185}
\DefMacro{guavaForceGCAssertQuery}{75}
\DefMacro{guavaCacheInstanceInfoAvg}{2697}
\DefMacro{guavaCacheInstanceInfoNumRuns}{50}
\DefMacro{guavaCacheInstanceInfoStdev}{102}
\DefMacro{guavaCacheInstanceInfoAssertQuery}{78}
\DefMacro{guavaAntlrWhitelistAvg}{218}
\DefMacro{guavaAntlrWhitelistNumRuns}{50}
\DefMacro{guavaAntlrWhitelistStdev}{6}
\DefMacro{guavaAntlrNoneAvg}{217}
\DefMacro{guavaAntlrNoneNumRuns}{50}
\DefMacro{guavaAntlrNoneStdev}{6}
\DefMacro{guavaAntlrForceGCAvg}{130}
\DefMacro{guavaAntlrForceGCNumRuns}{50}
\DefMacro{guavaAntlrForceGCStdev}{8}
\DefMacro{guavaAllNumObjects}{49969}
\DefMacro{guavaWhitelistNumObjects}{291537}
\DefMacro{guavaForceGCNumObjects}{49969}
\DefMacro{guavaAntlrWhitelistNumObjects}{283106}
\DefMacro{guavaAntlrNoneNumObjects}{283089}
\DefMacro{guavaAntlrForceGCNumObjects}{48663}
\DefMacro{guavaAntlrNumObjectsReduction}{82.81}
\DefMacro{guavaAntlrFGCNeoAll}{95}

\DefMacro{jfreeChartTraversePassedObjectsAvg}{1824}
\DefMacro{jfreeChartTraversePassedObjectsNumRuns}{50}
\DefMacro{jfreeChartTraversePassedObjectsStdev}{199}
\DefMacro{jfreeChartTraversePassedObjectsAssertQuery}{196}
\DefMacro{jfreeChartAllAvg}{1548}
\DefMacro{jfreeChartAllNumRuns}{50}
\DefMacro{jfreeChartAllStdev}{179}
\DefMacro{jfreeChartAllAssertQuery}{186}
\DefMacro{jfreeChartWhitelistAvg}{1811}
\DefMacro{jfreeChartWhitelistNumRuns}{50}
\DefMacro{jfreeChartWhitelistStdev}{221}
\DefMacro{jfreeChartWhitelistAssertQuery}{199}
\DefMacro{jfreeChartInstanceInfoReallocAvg}{1778}
\DefMacro{jfreeChartInstanceInfoReallocNumRuns}{50}
\DefMacro{jfreeChartInstanceInfoReallocStdev}{202}
\DefMacro{jfreeChartInstanceInfoReallocAssertQuery}{197}
\DefMacro{jfreeChartNoneAvg}{383694}
\DefMacro{jfreeChartNoneNumRuns}{2}
\DefMacro{jfreeChartNoneAssertQuery}{91}
\DefMacro{jfreeChartForceGCAvg}{1625}
\DefMacro{jfreeChartForceGCNumRuns}{50}
\DefMacro{jfreeChartForceGCStdev}{242}
\DefMacro{jfreeChartForceGCAssertQuery}{187}
\DefMacro{jfreeChartCacheInstanceInfoAvg}{1841}
\DefMacro{jfreeChartCacheInstanceInfoNumRuns}{50}
\DefMacro{jfreeChartCacheInstanceInfoStdev}{308}
\DefMacro{jfreeChartCacheInstanceInfoAssertQuery}{198}
\DefMacro{jfreeChartAntlrWhitelistAvg}{234}
\DefMacro{jfreeChartAntlrWhitelistNumRuns}{50}
\DefMacro{jfreeChartAntlrWhitelistStdev}{8}
\DefMacro{jfreeChartAntlrNoneAvg}{256}
\DefMacro{jfreeChartAntlrNoneNumRuns}{50}
\DefMacro{jfreeChartAntlrNoneStdev}{8}
\DefMacro{jfreeChartAntlrForceGCAvg}{134}
\DefMacro{jfreeChartAntlrForceGCNumRuns}{50}
\DefMacro{jfreeChartAntlrForceGCStdev}{12}
\DefMacro{jfreeChartWhitelistNumObjects}{2365782}
\DefMacro{jfreeChartForceGCNumObjects}{592317}
\DefMacro{jfreeChartAntlrWhitelistNumObjects}{327834}
\DefMacro{jfreeChartAntlrNoneNumObjects}{327834}
\DefMacro{jfreeChartAntlrForceGCNumObjects}{62435}
\DefMacro{jfreeChartAntlrNumObjectsReduction}{80.96}
\DefMacro{jfreeChartAntlrFGCNeoAll}{91}

\DefMacro{jsonJavaTraversePassedObjectsAvg}{1423}
\DefMacro{jsonJavaTraversePassedObjectsNumRuns}{50}
\DefMacro{jsonJavaTraversePassedObjectsStdev}{86}
\DefMacro{jsonJavaTraversePassedObjectsAssertQuery}{85}
\DefMacro{jsonJavaAllAvg}{1400}
\DefMacro{jsonJavaAllNumRuns}{50}
\DefMacro{jsonJavaAllStdev}{66}
\DefMacro{jsonJavaAllAssertQuery}{89}
\DefMacro{jsonJavaWhitelistAvg}{1456}
\DefMacro{jsonJavaWhitelistNumRuns}{50}
\DefMacro{jsonJavaWhitelistStdev}{76}
\DefMacro{jsonJavaWhitelistAssertQuery}{85}
\DefMacro{jsonJavaInstanceInfoReallocAvg}{1415}
\DefMacro{jsonJavaInstanceInfoReallocNumRuns}{50}
\DefMacro{jsonJavaInstanceInfoReallocStdev}{88}
\DefMacro{jsonJavaInstanceInfoReallocAssertQuery}{85}
\DefMacro{jsonJavaNoneAvg}{99798}
\DefMacro{jsonJavaNoneNumRuns}{2}
\DefMacro{jsonJavaNoneAssertQuery}{74}
\DefMacro{jsonJavaForceGCAvg}{1483}
\DefMacro{jsonJavaForceGCNumRuns}{50}
\DefMacro{jsonJavaForceGCStdev}{197}
\DefMacro{jsonJavaForceGCAssertQuery}{88}
\DefMacro{jsonJavaCacheInstanceInfoAvg}{1424}
\DefMacro{jsonJavaCacheInstanceInfoNumRuns}{50}
\DefMacro{jsonJavaCacheInstanceInfoStdev}{82}
\DefMacro{jsonJavaCacheInstanceInfoAssertQuery}{85}
\DefMacro{jsonJavaNaive}{\textbf{OOM}}
\DefMacro{jsonJavaNaiveAssertQuery}{\textbf{OOM}}
\DefMacro{jsonJavaAntlrWhitelistAvg}{155}
\DefMacro{jsonJavaAntlrWhitelistNumRuns}{50}
\DefMacro{jsonJavaAntlrWhitelistStdev}{4}
\DefMacro{jsonJavaAntlrNoneAvg}{155}
\DefMacro{jsonJavaAntlrNoneNumRuns}{50}
\DefMacro{jsonJavaAntlrNoneStdev}{4}
\DefMacro{jsonJavaAntlrForceGCAvg}{126}
\DefMacro{jsonJavaAntlrForceGCNumRuns}{50}
\DefMacro{jsonJavaAntlrForceGCStdev}{7}
\DefMacro{jsonJavaAllNumObjects}{48995}
\DefMacro{jsonJavaWhitelistNumObjects}{160024}
\DefMacro{jsonJavaForceGCNumObjects}{48995}
\DefMacro{jsonJavaAntlrWhitelistNumObjects}{154465}
\DefMacro{jsonJavaAntlrNoneNumObjects}{154466}
\DefMacro{jsonJavaAntlrForceGCNumObjects}{47706}
\DefMacro{jsonJavaAntlrNumObjectsReduction}{69.11}
\DefMacro{jsonJavaAntlrFGCNeoAll}{91}

\DefMacro{zipFourjTraversePassedObjectsAvg}{3548}
\DefMacro{zipFourjTraversePassedObjectsNumRuns}{50}
\DefMacro{zipFourjTraversePassedObjectsStdev}{680}
\DefMacro{zipFourjTraversePassedObjectsAssertQuery}{148}
\DefMacro{zipFourjAllAvg}{3157}
\DefMacro{zipFourjAllNumRuns}{50}
\DefMacro{zipFourjAllStdev}{417}
\DefMacro{zipFourjAllAssertQuery}{147}
\DefMacro{zipFourjWhitelistAvg}{3528}
\DefMacro{zipFourjWhitelistNumRuns}{50}
\DefMacro{zipFourjWhitelistStdev}{639}
\DefMacro{zipFourjWhitelistAssertQuery}{148}
\DefMacro{zipFourjInstanceInfoReallocAvg}{3460}
\DefMacro{zipFourjInstanceInfoReallocNumRuns}{50}
\DefMacro{zipFourjInstanceInfoReallocStdev}{483}
\DefMacro{zipFourjInstanceInfoReallocAssertQuery}{149}
\DefMacro{zipFourjNoneAvg}{79658}
\DefMacro{zipFourjNoneNumRuns}{2}
\DefMacro{zipFourjNoneAssertQuery}{78}
\DefMacro{zipFourjForceGCAvg}{3322}
\DefMacro{zipFourjForceGCNumRuns}{50}
\DefMacro{zipFourjForceGCStdev}{545}
\DefMacro{zipFourjForceGCAssertQuery}{147}
\DefMacro{zipFourjCacheInstanceInfoAvg}{3523}
\DefMacro{zipFourjCacheInstanceInfoNumRuns}{50}
\DefMacro{zipFourjCacheInstanceInfoStdev}{763}
\DefMacro{zipFourjCacheInstanceInfoAssertQuery}{147}
\DefMacro{zipFourjAntlrWhitelistAvg}{242}
\DefMacro{zipFourjAntlrWhitelistNumRuns}{50}
\DefMacro{zipFourjAntlrWhitelistStdev}{19}
\DefMacro{zipFourjAntlrNoneAvg}{244}
\DefMacro{zipFourjAntlrNoneNumRuns}{50}
\DefMacro{zipFourjAntlrNoneStdev}{17}
\DefMacro{zipFourjAntlrForceGCAvg}{174}
\DefMacro{zipFourjAntlrForceGCNumRuns}{50}
\DefMacro{zipFourjAntlrForceGCStdev}{15}
\DefMacro{zipFourjAllNumObjects}{618660}
\DefMacro{zipFourjWhitelistNumObjects}{193810}
\DefMacro{zipFourjForceGCNumObjects}{66300}
\DefMacro{zipFourjAntlrWhitelistNumObjects}{214930}
\DefMacro{zipFourjAntlrNoneNumObjects}{215603}
\DefMacro{zipFourjAntlrForceGCNumObjects}{73843}
\DefMacro{zipFourjAntlrNumObjectsReduction}{65.64}
\DefMacro{zipFourjAntlrFGCNeoAll}{94}

\DefMacro{tcasesTraversePassedObjectsAvg}{39713}
\DefMacro{tcasesTraversePassedObjectsNumRuns}{50}
\DefMacro{tcasesTraversePassedObjectsStdev}{1836}
\DefMacro{tcasesTraversePassedObjectsAssertQuery}{212}
\DefMacro{tcasesAllAvg}{2068}
\DefMacro{tcasesAllNumRuns}{50}
\DefMacro{tcasesAllStdev}{55}
\DefMacro{tcasesAllAssertQuery}{63}
\DefMacro{tcasesWhitelistAvg}{39736}
\DefMacro{tcasesWhitelistNumRuns}{50}
\DefMacro{tcasesWhitelistStdev}{1570}
\DefMacro{tcasesWhitelistAssertQuery}{218}
\DefMacro{tcasesInstanceInfoReallocAvg}{39352}
\DefMacro{tcasesInstanceInfoReallocNumRuns}{50}
\DefMacro{tcasesInstanceInfoReallocStdev}{1543}
\DefMacro{tcasesInstanceInfoReallocAssertQuery}{214}
\DefMacro{tcasesNoneAvg}{323006}
\DefMacro{tcasesNoneNumRuns}{2}
\DefMacro{tcasesNoneAssertQuery}{222}
\DefMacro{tcasesForceGCAvg}{2120}
\DefMacro{tcasesForceGCNumRuns}{50}
\DefMacro{tcasesForceGCStdev}{60}
\DefMacro{tcasesForceGCAssertQuery}{67}
\DefMacro{tcasesCacheInstanceInfoAvg}{39497}
\DefMacro{tcasesCacheInstanceInfoNumRuns}{50}
\DefMacro{tcasesCacheInstanceInfoStdev}{1555}
\DefMacro{tcasesCacheInstanceInfoAssertQuery}{217}
\DefMacro{tcasesAntlrWhitelistAvg}{796}
\DefMacro{tcasesAntlrWhitelistNumRuns}{50}
\DefMacro{tcasesAntlrWhitelistStdev}{66}
\DefMacro{tcasesAntlrNoneAvg}{768}
\DefMacro{tcasesAntlrNoneNumRuns}{50}
\DefMacro{tcasesAntlrNoneStdev}{61}
\DefMacro{tcasesAntlrForceGCAvg}{199}
\DefMacro{tcasesAntlrForceGCNumRuns}{50}
\DefMacro{tcasesAntlrForceGCStdev}{14}
\DefMacro{tcasesAllNumObjects}{72940}
\DefMacro{tcasesWhitelistNumObjects}{344539}
\DefMacro{tcasesForceGCNumObjects}{72940}
\DefMacro{tcasesAntlrWhitelistNumObjects}{881221}
\DefMacro{tcasesAntlrNoneNumObjects}{875154}
\DefMacro{tcasesAntlrForceGCNumObjects}{76464}
\DefMacro{tcasesAntlrNumObjectsReduction}{91.32}
\DefMacro{tcasesAntlrFGCNeoAll}{90}

\DefMacro{assertjdbVanillaAssertion}{0.34}
\DefMacro{commonscliVanillaAssertion}{0.009}
\DefMacro{commonscsvVanillaAssertion}{0.013}
\DefMacro{commonslangVanillaAssertion}{0.014}
\DefMacro{guavaVanillaAssertion}{0.138}
\DefMacro{jfreechartVanillaAssertion}{2.797}
\DefMacro{JSONjavaVanillaAssertion}{0.085}
\DefMacro{zip4jVanillaAssertion}{0.034}

\DefMacro{TableLocNocLoc}{\textbf{LOC}}
\DefMacro{TableLocNocNoc}{\textbf{NOC}}
\DefMacro{TableLocNocOgo}{\textbf{OGO}}
\DefMacro{TableLocNocImp}{\textbf{Imperative}}
\DefMacro{TableLocNocClass}{\textbf{Class}}
\DefMacro{TableLocNocAvg}{\textbf{Avg.}}
\DefMacro{TableLocNocSum}{$\pmb{\sum}$}
\DefMacro{TableLocNocCaption}{Comparison of LOC(lines of code) and NOC(number of characters) between OGO and imperative implementation of selected instance methods.}
\DefMacro{TableLocNocGuava}{Guava}
\DefMacro{TableLocNocArrayTable}{ArrayTable}
\DefMacro{TableLocNocArrayTableOgoNoc}{315}
\DefMacro{TableLocNocArrayTableImpNoc}{1126}
\DefMacro{TableLocNocArrayTableOgoLoc}{9}
\DefMacro{TableLocNocArrayTableImpLoc}{35}
\DefMacro{TableLocNocHashBiMap}{HashBiMap}
\DefMacro{TableLocNocHashBiMapOgoNoc}{226}
\DefMacro{TableLocNocHashBiMapImpNoc}{852}
\DefMacro{TableLocNocHashBiMapOgoLoc}{6}
\DefMacro{TableLocNocHashBiMapImpLoc}{26}
\DefMacro{TableLocNocLinkedListMultiMap}{LinkedListMultiMap}
\DefMacro{TableLocNocLinkedListMultiMapOgoNoc}{202}
\DefMacro{TableLocNocLinkedListMultiMapImpNoc}{1322}
\DefMacro{TableLocNocLinkedListMultiMapOgoLoc}{6}
\DefMacro{TableLocNocLinkedListMultiMapImpLoc}{48}
\DefMacro{TableLocNocJUNG}{JUNG}
\DefMacro{TableLocNocSparseGraph}{SparseGraph}
\DefMacro{TableLocNocSparseGraphOgoNoc}{303}
\DefMacro{TableLocNocSparseGraphImpNoc}{1208}
\DefMacro{TableLocNocSparseGraphOgoLoc}{7}
\DefMacro{TableLocNocSparseGraphImpLoc}{38}
\DefMacro{TableLocNocUndirectedSparseGraph}{UndirectedSparseGraph}
\DefMacro{TableLocNocUndirectedSparseGraphOgoNoc}{281}
\DefMacro{TableLocNocUndirectedSparseGraphImpNoc}{1066}
\DefMacro{TableLocNocUndirectedSparseGraphOgoLoc}{6}
\DefMacro{TableLocNocUndirectedSparseGraphImpLoc}{36}
\DefMacro{TableLocNocDirectedSparseGraph}{DirectedSparseGraph}
\DefMacro{TableLocNocDirectedSparseGraphOgoNoc}{281}
\DefMacro{TableLocNocDirectedSparseGraphImpNoc}{1078}
\DefMacro{TableLocNocDirectedSparseGraphOgoLoc}{6}
\DefMacro{TableLocNocDirectedSparseGraphImpLoc}{36}
\DefMacro{TableLocNocJCF}{JCF}
\DefMacro{TableLocNocArrayList}{ArrayList}
\DefMacro{TableLocNocArrayListOgoNoc}{212}
\DefMacro{TableLocNocArrayListImpNoc}{426}
\DefMacro{TableLocNocArrayListOgoLoc}{6}
\DefMacro{TableLocNocArrayListImpLoc}{25}
\DefMacro{TableLocNocHashMap}{HashMap}
\DefMacro{TableLocNocHashMapOgoNoc}{214}
\DefMacro{TableLocNocHashMapImpNoc}{1073}
\DefMacro{TableLocNocHashMapOgoLoc}{6}
\DefMacro{TableLocNocHashMapImpLoc}{37}
\DefMacro{TableLocNocLinkedList}{LinkedList}
\DefMacro{TableLocNocLinkedListOgoNoc}{226}
\DefMacro{TableLocNocLinkedListImpNoc}{374}
\DefMacro{TableLocNocLinkedListOgoLoc}{6}
\DefMacro{TableLocNocLinkedListImpLoc}{21}
\DefMacro{TableLocNocArrayDeque}{ArrayDeque}
\DefMacro{TableLocNocArrayDequeOgoNoc}{99}
\DefMacro{TableLocNocArrayDequeImpNoc}{326}
\DefMacro{TableLocNocArrayDequeOgoLoc}{3}
\DefMacro{TableLocNocArrayDequeImpLoc}{15}
\DefMacro{TableLocNocVector}{Vector}
\DefMacro{TableLocNocVectorOgoNoc}{213}
\DefMacro{TableLocNocVectorImpNoc}{413}
\DefMacro{TableLocNocVectorOgoLoc}{6}
\DefMacro{TableLocNocVectorImpLoc}{20}
\DefMacro{TableLocNocAvgValueOgoLoc}{6}
\DefMacro{TableLocNocSumValueOgoLoc}{67}
\DefMacro{TableLocNocAvgValueOgoNoc}{234}
\DefMacro{TableLocNocSumValueOgoNoc}{2572}
\DefMacro{TableLocNocAvgValueImpLoc}{31}
\DefMacro{TableLocNocSumValueImpLoc}{337}
\DefMacro{TableLocNocAvgValueImpNoc}{842}
\DefMacro{TableLocNocSumValueImpNoc}{9264}

\DefMacro{TableDataStructureEvalClass}{\textbf{Class}}
\DefMacro{TableDataStructureEvalTests}{\textbf{\#Tests}}
\DefMacro{TableDataStructureEvalWL}{\textbf{+WL}}
\DefMacro{TableDataStructureEvalFRO}{\textbf{+FRO}}
\DefMacro{TableDataStructureEvalFGC}{\textbf{+FGC}}
\DefMacro{TableDataStructureEvalFROFGC}{\textbf{+FRO+FGC}}
\DefMacro{TableDataStructureEvalOGONeo}{\textbf{OGO$^{Neo}$}}
\DefMacro{TableDataStructureEvalOGOMem}{\textbf{OGO$^{Mem}$}}
\DefMacro{TableDataStructureEvalCaption}{\TableLibraryMethodsPerformanceCaption}
\DefMacro{TableDataStructureEvalVector}{Vector}
\DefMacro{TableDataStructureEvalVectorTwoHundred}{200}
\DefMacro{TableDataStructureEvalNeoVectorTwoHundredWLAvg}{1125}
\DefMacro{TableDataStructureEvalMemVectorTwoHundredWLAvg}{873}
\DefMacro{TableDataStructureEvalNeoVectorTwoHundredWLFGCAvg}{794}
\DefMacro{TableDataStructureEvalMemVectorTwoHundredWLFGCAvg}{316}
\DefMacro{TableDataStructureEvalNeoVectorTwoHundredWLFROAvg}{1254}
\DefMacro{TableDataStructureEvalMemVectorTwoHundredWLFROAvg}{980}
\DefMacro{TableDataStructureEvalNeoVectorTwoHundredWLFROFGCAvg}{823}
\DefMacro{TableDataStructureEvalMemVectorTwoHundredWLFROFGCAvg}{290}
\DefMacro{TableDataStructureEvalUndirectedSparseGraph}{UndirectedSparseGraph}
\DefMacro{TableDataStructureEvalUndirectedSparseGraphTwoHundred}{200}
\DefMacro{TableDataStructureEvalNeoUndirectedSparseGraphTwoHundredWLAvg}{7808}
\DefMacro{TableDataStructureEvalMemUndirectedSparseGraphTwoHundredWLAvg}{1032}
\DefMacro{TableDataStructureEvalNeoUndirectedSparseGraphTwoHundredWLFGCAvg}{2287}
\DefMacro{TableDataStructureEvalMemUndirectedSparseGraphTwoHundredWLFGCAvg}{186}
\DefMacro{TableDataStructureEvalNeoUndirectedSparseGraphTwoHundredWLFROAvg}{24986}
\DefMacro{TableDataStructureEvalMemUndirectedSparseGraphTwoHundredWLFROAvg}{1045}
\DefMacro{TableDataStructureEvalNeoUndirectedSparseGraphTwoHundredWLFROFGCAvg}{8499}
\DefMacro{TableDataStructureEvalMemUndirectedSparseGraphTwoHundredWLFROFGCAvg}{293}
\DefMacro{TableDataStructureEvalHashBiMap}{HashBiMap}
\DefMacro{TableDataStructureEvalHashBiMapTwoHundred}{200}
\DefMacro{TableDataStructureEvalNeoHashBiMapTwoHundredWLAvg}{2470}
\DefMacro{TableDataStructureEvalMemHashBiMapTwoHundredWLAvg}{519}
\DefMacro{TableDataStructureEvalNeoHashBiMapTwoHundredWLFGCAvg}{856}
\DefMacro{TableDataStructureEvalMemHashBiMapTwoHundredWLFGCAvg}{196}
\DefMacro{TableDataStructureEvalNeoHashBiMapTwoHundredWLFROAvg}{2635}
\DefMacro{TableDataStructureEvalMemHashBiMapTwoHundredWLFROAvg}{526}
\DefMacro{TableDataStructureEvalNeoHashBiMapTwoHundredWLFROFGCAvg}{782}
\DefMacro{TableDataStructureEvalMemHashBiMapTwoHundredWLFROFGCAvg}{179}
\DefMacro{TableDataStructureEvalHashMap}{HashMap}
\DefMacro{TableDataStructureEvalHashMapTwoHundred}{200}
\DefMacro{TableDataStructureEvalNeoHashMapTwoHundredWLAvg}{2693}
\DefMacro{TableDataStructureEvalMemHashMapTwoHundredWLAvg}{1166}
\DefMacro{TableDataStructureEvalNeoHashMapTwoHundredWLFGCAvg}{1771}
\DefMacro{TableDataStructureEvalMemHashMapTwoHundredWLFGCAvg}{315}
\DefMacro{TableDataStructureEvalNeoHashMapTwoHundredWLFROAvg}{2997}
\DefMacro{TableDataStructureEvalMemHashMapTwoHundredWLFROAvg}{1326}
\DefMacro{TableDataStructureEvalNeoHashMapTwoHundredWLFROFGCAvg}{1529}
\DefMacro{TableDataStructureEvalMemHashMapTwoHundredWLFROFGCAvg}{289}
\DefMacro{TableDataStructureEvalSparseGraph}{SparseGraph}
\DefMacro{TableDataStructureEvalSparseGraphTwoHundred}{200}
\DefMacro{TableDataStructureEvalNeoSparseGraphTwoHundredWLAvg}{13533}
\DefMacro{TableDataStructureEvalMemSparseGraphTwoHundredWLAvg}{1363}
\DefMacro{TableDataStructureEvalNeoSparseGraphTwoHundredWLFGCAvg}{2954}
\DefMacro{TableDataStructureEvalMemSparseGraphTwoHundredWLFGCAvg}{188}
\DefMacro{TableDataStructureEvalNeoSparseGraphTwoHundredWLFROAvg}{16945}
\DefMacro{TableDataStructureEvalMemSparseGraphTwoHundredWLFROAvg}{1444}
\DefMacro{TableDataStructureEvalNeoSparseGraphTwoHundredWLFROFGCAvg}{2666}
\DefMacro{TableDataStructureEvalMemSparseGraphTwoHundredWLFROFGCAvg}{167}
\DefMacro{TableDataStructureEvalArrayDeque}{ArrayDeque}
\DefMacro{TableDataStructureEvalArrayDequeTwoHundred}{200}
\DefMacro{TableDataStructureEvalNeoArrayDequeTwoHundredWLAvg}{1563}
\DefMacro{TableDataStructureEvalMemArrayDequeTwoHundredWLAvg}{1151}
\DefMacro{TableDataStructureEvalNeoArrayDequeTwoHundredWLFGCAvg}{1139}
\DefMacro{TableDataStructureEvalMemArrayDequeTwoHundredWLFGCAvg}{320}
\DefMacro{TableDataStructureEvalNeoArrayDequeTwoHundredWLFROAvg}{1442}
\DefMacro{TableDataStructureEvalMemArrayDequeTwoHundredWLFROAvg}{1104}
\DefMacro{TableDataStructureEvalNeoArrayDequeTwoHundredWLFROFGCAvg}{1047}
\DefMacro{TableDataStructureEvalMemArrayDequeTwoHundredWLFROFGCAvg}{292}
\DefMacro{TableDataStructureEvalArrayList}{ArrayList}
\DefMacro{TableDataStructureEvalArrayListTwoHundred}{200}
\DefMacro{TableDataStructureEvalNeoArrayListTwoHundredWLAvg}{1391}
\DefMacro{TableDataStructureEvalMemArrayListTwoHundredWLAvg}{1080}
\DefMacro{TableDataStructureEvalNeoArrayListTwoHundredWLFGCAvg}{1152}
\DefMacro{TableDataStructureEvalMemArrayListTwoHundredWLFGCAvg}{312}
\DefMacro{TableDataStructureEvalNeoArrayListTwoHundredWLFROAvg}{1476}
\DefMacro{TableDataStructureEvalMemArrayListTwoHundredWLFROAvg}{1128}
\DefMacro{TableDataStructureEvalNeoArrayListTwoHundredWLFROFGCAvg}{1131}
\DefMacro{TableDataStructureEvalMemArrayListTwoHundredWLFROFGCAvg}{292}
\DefMacro{TableDataStructureEvalArrayTable}{ArrayTable}
\DefMacro{TableDataStructureEvalArrayTableTwoHundred}{200}
\DefMacro{TableDataStructureEvalNeoArrayTableTwoHundredWLAvg}{5278}
\DefMacro{TableDataStructureEvalMemArrayTableTwoHundredWLAvg}{624}
\DefMacro{TableDataStructureEvalNeoArrayTableTwoHundredWLFGCAvg}{1313}
\DefMacro{TableDataStructureEvalMemArrayTableTwoHundredWLFGCAvg}{193}
\DefMacro{TableDataStructureEvalNeoArrayTableTwoHundredWLFROAvg}{5786}
\DefMacro{TableDataStructureEvalMemArrayTableTwoHundredWLFROAvg}{652}
\DefMacro{TableDataStructureEvalNeoArrayTableTwoHundredWLFROFGCAvg}{1332}
\DefMacro{TableDataStructureEvalMemArrayTableTwoHundredWLFROFGCAvg}{178}
\DefMacro{TableDataStructureEvalLinkedList}{LinkedList}
\DefMacro{TableDataStructureEvalLinkedListTwoHundred}{200}
\DefMacro{TableDataStructureEvalNeoLinkedListTwoHundredWLAvg}{838}
\DefMacro{TableDataStructureEvalMemLinkedListTwoHundredWLAvg}{492}
\DefMacro{TableDataStructureEvalNeoLinkedListTwoHundredWLFGCAvg}{354}
\DefMacro{TableDataStructureEvalMemLinkedListTwoHundredWLFGCAvg}{190}
\DefMacro{TableDataStructureEvalNeoLinkedListTwoHundredWLFROAvg}{3101}
\DefMacro{TableDataStructureEvalMemLinkedListTwoHundredWLFROAvg}{638}
\DefMacro{TableDataStructureEvalNeoLinkedListTwoHundredWLFROFGCAvg}{827}
\DefMacro{TableDataStructureEvalMemLinkedListTwoHundredWLFROFGCAvg}{257}
\DefMacro{TableDataStructureEvalDirectedSparseGraph}{DirectedSparseGraph}
\DefMacro{TableDataStructureEvalDirectedSparseGraphTwoHundred}{200}
\DefMacro{TableDataStructureEvalNeoDirectedSparseGraphTwoHundredWLAvg}{3371}
\DefMacro{TableDataStructureEvalMemDirectedSparseGraphTwoHundredWLAvg}{1107}
\DefMacro{TableDataStructureEvalNeoDirectedSparseGraphTwoHundredWLFGCAvg}{3160}
\DefMacro{TableDataStructureEvalMemDirectedSparseGraphTwoHundredWLFGCAvg}{188}
\DefMacro{TableDataStructureEvalNeoDirectedSparseGraphTwoHundredWLFROAvg}{10549}
\DefMacro{TableDataStructureEvalMemDirectedSparseGraphTwoHundredWLFROAvg}{1215}
\DefMacro{TableDataStructureEvalNeoDirectedSparseGraphTwoHundredWLFROFGCAvg}{3240}
\DefMacro{TableDataStructureEvalMemDirectedSparseGraphTwoHundredWLFROFGCAvg}{177}
\DefMacro{TableDataStructureEvalNeoWLAvg}{4007}
\DefMacro{TableDataStructureEvalNeoWLTotal}{40070}
\DefMacro{TableDataStructureEvalNeoWLFROAvg}{1578}
\DefMacro{TableDataStructureEvalNeoWLFROTotal}{15780}
\DefMacro{TableDataStructureEvalNeoWLFGCAvg}{7117}
\DefMacro{TableDataStructureEvalNeoWLFGCTotal}{71171}
\DefMacro{TableDataStructureEvalNeoWLFROFGCAvg}{2188}
\DefMacro{TableDataStructureEvalNeoWLFROFGCTotal}{21876}
\DefMacro{TableDataStructureEvalMemWLAvg}{941}
\DefMacro{TableDataStructureEvalMemWLTotal}{9407}
\DefMacro{TableDataStructureEvalMemWLFROAvg}{240}
\DefMacro{TableDataStructureEvalMemWLFROTotal}{2404}
\DefMacro{TableDataStructureEvalMemWLFGCAvg}{1006}
\DefMacro{TableDataStructureEvalMemWLFGCTotal}{10058}
\DefMacro{TableDataStructureEvalMemWLFROFGCAvg}{241}
\DefMacro{TableDataStructureEvalMemWLFROFGCTotal}{2414}

\usepackage{mathpartir}

\newcommand{\Nods}{\ensuremath{\mathcal{N}}}
\newcommand{\Rels}{\ensuremath{\mathcal{R}}}
\newcommand{\Labs}{\ensuremath{\mathcal{L}}}
\newcommand{\Props}{\ensuremath{\mathcal{P}}}

\newcommand*{\tC}{\mathtt{C}}
\newcommand*{\tCs}{\overline{\tC}}
\newcommand*{\tD}{\mathtt{D}}
\newcommand*{\tM}{\mathtt{M}}
\newcommand*{\tMs}{\overline{\mathtt{M}}}
\newcommand*{\tm}{\mathtt{m}}
\newcommand*{\tK}{\mathtt{K}}
\newcommand*{\tL}{\mathtt{L}}
\newcommand*{\tf}{\mathtt{f}}
\newcommand*{\tfs}{\overline{\mathtt{f}}}
\newcommand*{\tclass}{\mathtt{class}}
\newcommand*{\textends}{\mathtt{extends}}
\newcommand*{\tsuper}{\mathtt{super}}
\newcommand*{\tthis}{\mathtt{this}}
\newcommand*{\treturn}{\mathtt{return}}
\newcommand*{\tnew}{\mathtt{new}}
\newcommand*{\te}{\mathtt{e}}
\newcommand*{\tc}{\mathtt{c}}

\newcommand*{\tx}{\mathtt{x}}
\newcommand*{\ty}{\mathtt{y}}
\newcommand*{\txs}{\overline{\tx}}
\newcommand*{\tys}{\overline{\ty}}

\newcommand{\classd}[2]{\ensuremath{
    \tclass~#1~\textends~#2~\{\overline{#1}~\tfs;~\tK~\tMs\}
}}
\newcommand{\constrd}{\ensuremath{
    \tC(\tCs~\txs)\{\tsuper(\txs);~\tthis.\tfs=\txs;\}
}}
\newcommand{\methodd}{\ensuremath{
    \tC ~ \tm(\tCs ~\txs)\{ \te \}
}}

\newcommand*{\ifs}{\mathit{fields}}
\newcommand*{\imt}{\mathit{mtype}}
\newcommand*{\imb}{\mathit{mbody}}

\newcommand*{\step}[3]{#2 \diamond #1 \longrightarrow #3}
\newcommand*{\stepM}[3]{#2 \diamond #1 \longrightarrow^* #3}

\newcommand*{\fs}{\mathit{mkFields}}
\newcommand*{\local}{\mathit{Local}}

\newcommand{\NumProjects}{10\xspace}
\newcommand{\NumAssertions}{230\xspace}

\definecolor{gray}{RGB}{211,211,211}
\definecolor{purple2}{RGB}{140, 59, 244}
\definecolor{armygreen}{rgb}{0.29, 0.5, 0.13}
\definecolor{britishracinggreen}{rgb}{0.0, 0.26, 0.15}
\newcommand{\jbasicstyle}{\small\sffamily} 

\newcommand{\jnumberstyle}{\scriptsize}

\lstdefinelanguage{pseudo}
{
  morekeywords={},
  keywordstyle=\bfseries,
  lineskip=-0.1em,
  numbers=left, 
  numberstyle=\jnumberstyle,
  numbersep=4pt,
  basicstyle=\jbasicstyle,
  breaklines=true,
  breakautoindent=true,
  tabsize=2,
  columns=fullflexible,
  morecomment=*[l][\textsl]{//},
  mathescape=true,
  xleftmargin=10pt,
}

\lstdefinelanguage{todo-comment}
{
  morekeywords={},
  keywordstyle=\bfseries,
  lineskip=-0.1em,
  numbers=none,
  basicstyle=\jbasicstyle,
  breaklines=true,
  breakautoindent=true,
  tabsize=2,
  columns=fullflexible,
  morecomment=*[l][\textsl]{//},
  mathescape=true,
  xleftmargin=-10pt,
}

\lstdefinelanguage{java-pretty}
{
  language=java,
  keywords=[2]{join, union, closure, rclosure, all, some, one, lone, none, in, not\_in, sub, old, intersect, filter, equals, range},
  keywordstyle=[2]\color{blue},
  keywords=[3]{invariant, specField, specCase, field, freshObject, freshObjects, modifies, ensures, requires},
  keywordstyle=[3]\color{purple}\textbf,
  numbers=left,
  numbersep=2pt,
  basicstyle=\footnotesize\ttfamily,
  numberstyle=\scriptsize,
  commentstyle=\color{darkgray},
  breaklines=true,
  columns=fullflexible,
  xleftmargin=8pt,
  xrightmargin=-10pt,
  showstringspaces=false
}

\lstdefinelanguage{small-java-pretty}
{
  language=java,
  numbers=left,
  numbersep=2pt,
  basicstyle=\scriptsize\ttfamily,
  numberstyle=\tiny,
  commentstyle=\color{brown},
  stringstyle=\color{britishracinggreen},
  keywordstyle=[2]\color{blue},
  keywordstyle=[3]\color{britishracinggreen}\textbf,
  morekeywords = [2]{query, queryBool, queryLong, queryInt, queryDouble, queryObject},
  morekeywords = [3]{MATCH, CREATE, MERGE, RETURN, OPTIONAL, WHERE},    
  breaklines=true,
  columns=fullflexible,
  xleftmargin=8pt,
  xrightmargin=0pt,
  showstringspaces=false,
  morestring=*[d]{"},
  escapeinside={!<}{>!}
}

\lstdefinelanguage{positional-argument-table}
{
  language=java,
  numbers=none,
  basicstyle=\scriptsize\ttfamily,
  numberstyle=\tiny,
  commentstyle=\color{brown},
  stringstyle=\color{britishracinggreen},
  keywordstyle=[2]\color{blue},
  morekeywords = [2]{query, queryBool, queryLong, queryInt, queryDouble, queryObject},  
  breaklines=true,
  columns=fullflexible,
  xleftmargin=1pt,
  aboveskip=-1.3ex,
  belowskip=-3.5ex,
  showstringspaces=false,
  morestring=*[d]{"},
  escapeinside={!<}{>!}
}

\lstdefinelanguage{java-ogo}
{
  language=java,
  numbers=left,
  numbersep=2pt,
  basicstyle=\scriptsize\ttfamily,
  numberstyle=\tiny,
  commentstyle=\color{darkgray},
  stringstyle=\color{britishracinggreen},
  keywordstyle=\color{blue},
  breaklines=true,
  columns=fullflexible,
  xleftmargin=8pt,
  xrightmargin=-10pt,
  showstringspaces=false
}

\lstdefinelanguage{bytecode}
{
  language=java,
  keywords=[2]{aload_0, getfield, invokedynamic, invokestatic, areturn, invokevirtual},
  keywordstyle=[2]\textbf,
  numbers=left,
  basicstyle=\footnotesize\ttfamily,
  numberstyle=\scriptsize,
  breaklines=true,
  columns=fullflexible,
  xleftmargin=8pt,
  showstringspaces=false,
  escapechar={!},
}

\newcommand{\FigCreateDSCaption}{Creating and querying instances and relations between instances using \Tool. Graphs above queries visualize the results of those queries. (a) The \Code{CREATE} clause is first used to create instances of \Code{BinaryTree} and \Code{Node} followed by which a \Code{MERGE} clause is used to create relations between the created instances. The reference to the created \Code{BinaryTree} instance (reachable from all instances created by the \Code{CREATE} clause under transitive closure) is returned at the end of the query to prevent the created instances from being garbage collected. (b) The query pattern is undirected and unconstrained in terms of the instance \Code{type} being matched and their distance from \textbf{n} (\Italicise{root}) and hence all reachable objects are returned. This is particularly useful for identifying object confinement issues. (c) The query pattern is directed and constrained to \Code{Node} instances reachable from \textbf{n} through fields \Code{left} or \Code{right} that are exactly 2 hops away.}
\newcommand{\FigCreateDSInstantiateCaption}{Creating and connecting \Code{Node} and \Code{BinaryTree} instances.}
\newcommand{\FigCreateDSMatchGenericReachabilityCaption}{Finding all objects reachable from an object.}
\newcommand{\FigCreateDSMatchSpecificReachabilityCaption}{Checking existence of a specific node.}

\newcommand{\FigKoratBTreeDefCaption}{\Code{BinaryTree} class definitions.}

\newcommand{\FigJavaDSCaption}{Methods from \Java collections framework using \Tool. (a) The nested static class \Code{Node} stores a key-value pair and the reference field \Code{table} stores all the entries in the map. (b) The imperative implementation of \Code{containsKey} uses the \Code{getNode} method. (c) \Tool (\ToolMemory) can also be used to invoke instance methods as shown in the \Code{WHERE} clause.}
\newcommand{\FigJavaDSDefCaption}{Snippet of \Code{java.util.HashMap} class definition.}
\newcommand{\FigJavaDSImperativeCaption}{Imperative implementation of \Code{HashMap}'s \Code{containsKey} method.}
\newcommand{\FigJavaDSOGOCaption}{\Tool implementation of \Code{HashMap}'s \Code{containsKey} method.}

\newcommand{\FigOgoApiCaption}{\Tool API available via the \Code{\Tool} class. Bounded queries (lines~\ref{line:bounded_query}, \ref{line:bounded_query_object}, \ref{line:other_query_begin} and \ref{line:bounded_query_long}) contain an additional root argument that constraints the query execution to a subgraph (limited to only objects reachable from the root argument under transitive closure) of the JVM heap object graph.}

\newcommand{\FigTranslationCaption}{An example class, its test class and the corresponding object graph shown as a property graph. (c) instances of \CodeIn{Node}, \CodeIn{BinaryTree} and \CodeIn{java.lang.Class} are shown colored black, blue and grey. The local
variable references are shown as dashed edges. Reference fields
(\CodeIn{left,right}) are mapped as node relations whereas primitive(
\CodeIn{value,size}) and String fields are mapped as node properties.}
\newcommand{\FigTranslationFigCaption}{Graph database at $\ast$ POINT $\ast$.}
\newcommand{\FigTranslationCodeCaption}{An example test class.}
\newcommand{\FigOGOStepBreakdownCaption}{Average time taken by major steps of \ToolNeo with the introduced optimizations.}
\newcommand{\TablePositionalArgumentsCaption}{Positional arguments supported by \Tool in the query formatting string.\vspace{-8pt}}
\newcommand{\TableLibraryMethodsPerformanceCaption}{Comparison of average execution time (ms) for instance methods implemented using OGO for random workloads.\vspace{-10pt}}

\usepackage{multicol}
\usepackage{algorithmic}
\usepackage{graphicx}
\usepackage{textcomp}
\usepackage{xcolor}
\usepackage{tkz-euclide}
\usepackage{comment}
\usepackage{framed}
\usepackage[frozencache]{minted}
\usepackage{booktabs}
\usepackage{caption,subcaption}
\usepackage{arydshln}
\usepackage{lineno}
\usepackage{tikz}
\usepackage{multirow}
\usepackage{tabularx}
\usepackage{xurl}
\usepackage{soul}
\usepackage{epstopdf}
\usepackage[thinlines]{easytable}
\usetikzlibrary{calc,positioning}
\usepackage{svg}
\def\BibTeX{{\rm B\kern-.05em{\sc i\kern-.025em b}\kern-.08em
    T\kern-.1667em\lower.7ex\hbox{E}\kern-.125emX}}
\usepackage{url}

\makeatletter

\newcommand*{\tabminted@finalstrut}[1]{%
  \ifdim\prevdepth>0pt
    \ifdim\dp#1>\prevdepth
      \vskip\dimexpr(\dp#1)-\prevdepth\relax
    \fi
  \else
    \vskip\dimexpr(\dp#1)\relax
  \fi
}
\newcommand*{\@tabmintedend}{%
  \let\@finalstrut\tabminted@finalstrut
}
\makeatother

\copyrightyear{2024}
\acmYear{2024}
\setcopyright{rightsretained}
\acmConference[ICSE '24]{2024 IEEE/ACM 46th International Conference on Software Engineering}{April 14--20, 2024}{Lisbon, Portugal} \acmBooktitle{2024 IEEE/ACM 46th International Conference on Software Engineering (ICSE '24), April 14--20, 2024, Lisbon, Portugal}\acmDOI{10.1145/3597503.3623319} \acmISBN{979-8-4007-0217-4/24/04}

\begin{document}

\setlength\dashlinedash{0.1pt}
\setlength\dashlinegap{0.1pt}
\setlength\arrayrulewidth{0.3pt}

\captionsetup[subfloat]{
  width=0.45\textwidth,
  aboveskip=2pt,
  belowskip=2pt,
  justification=raggedright
}

\captionsetup[figure, figure*]{
  aboveskip=2pt,
  belowskip=1pt
}


\title{Object Graph Programming}



\author{Aditya Thimmaiah}
\affiliation{%
  \institution{The University of Texas at Austin}
  \streetaddress{}
  \city{Austin} 
  \state{Texas}
  \country{USA}
  \postcode{78712}
}
\email{auditt@utexas.edu}

\author{Leonidas Lampropoulos}
\affiliation{%
  \institution{University of Maryland}
  \streetaddress{}
  \city{College Park} 
  \state{Maryland}
  \country{USA}
  \postcode{20742}
}
\email{leonidas@umd.edu}

\author{Christopher J. Rossbach}
\affiliation{%
  \institution{The University of Texas at Austin}
  \streetaddress{}
  \city{Austin} 
  \state{Texas}
  \country{USA} 
  \postcode{78712}
}
\email{rossbach@cs.utexas.edu}

\author{Milos Gligoric}
\affiliation{%
  \institution{The University of Texas at Austin}
  \streetaddress{}
  \city{Austin} 
  \state{Texas}
  \country{USA} 
  \postcode{78712}
}
\email{gligoric@utexas.edu}





\begin{abstract}
  We introduce \OGO (\Tool), which enables reading
  and modifying an object graph (i.e., the entire state of the object
  heap) via declarative queries.
  \Tool models the objects and their relations in the heap as an object
  graph thereby treating the heap as a graph database: each node in
  the graph is an object (e.g., an instance of a class or an instance
  of a metadata class) and each edge is a relation between objects
  (e.g., a field of one object references another object).
  We leverage \Cypher, the most popular query language for graph
  databases, as \Tool's query language.
  Unlike \LINQ, which uses collections (e.g., List) as a source of
  data, \Tool views the entire object graph as a single
  ``collection''.  \Tool is ideal for querying collections (just like
  \LINQ), introspecting the runtime system state (e.g., finding all
  instances of a given class or accessing fields via reflection), and
  writing assertions that have access to the entire program state.  We
  prototyped \Tool for Java in two ways: (a)~by translating an object
  graph into a Neo4j database on which we run \Cypher queries, and
  (b)~by implementing our own in-memory graph query engine that
  directly queries the object heap.
  We used \Tool to rewrite hundreds of statements in large open-source
  projects into \Tool queries.  We report our experience and
  performance of our prototypes.
\end{abstract}

\keywords{Object graph, graph database, query, \Cypher}
\maketitle

\pagestyle{plain}

\section{Introduction}
\label{sec:intro}

Declarative programming~\cite{kifer_declarative_programming}, focusing
on the \emph{what} rather than the \emph{how}, has grown into the
predominant way of programming in an increasing number of domains.
For instance, Structured Query Language (\SQL), a canonical example of
the declarative paradigm, is the primary query language for most
relational database management systems~\cite{Melton_SQL, Date_SQL,
  Jan_SQL, Begoli_SQL}.
At the same time, NoSQL databases have been gaining traction.  In
particular, the space of \emph{graph
  databases}~\cite{Angles_GraphDB,Jaroslav_GraphDB,Jouili_GraphDB,Jaroslav_GraphDB2}
is growing at a rapid pace,
as they have been shown to be a great fit for tasks such as fraud
detection, drug discovery~\cite{Soni_GraphDB_UseCase}, recommendation
engines, and data visualization~\cite{Kumar_GraphDB_UseCase,
  Consens_GraphDB_UseCase}.
Graph databases store data as property
graphs~\cite{Angles_PropertyGraphs, Hogan_PropertyGraphs,
  Maiolo_PropertyGraphs}, which emphasize relationships between data.

A \emph{property graph} (graph for short) contains nodes \Nods{} and
edges \Rels{} denoting relationships between nodes.  Each node is
assigned a \emph{label} \Labs{} and contains an arbitrary set of
\emph{properties} \Props: mappings from nodes to values.
Edges also have a label (sometimes called type in the literature) and
an arbitrary set of properties.
Querying, updating, and administering of such a graph is performed
with a \emph{graph query language}.
\Cypher~\cite{Francis_Cypher,Holzschuher_Cypher},
initially developed as part of the
Neo4j project~\cite{Neo4j}, is currently the most popular graph query
language~\cite{Seifer_Cypher_Popularity, Florian_Cypher_Popularity}.
\Cypher is a declarative language, in many ways similar to SQL, which
emphasizes simplicity and expressivity.  As an example, to get the
values of all nodes in a graph database we could run the following
query: \Code{match(n:\, Node)\; return\; n.val}.
Although graph databases have been used for various tasks, \emph{the
power of property graphs and graph query languages has yet to be used
to enhance developers' experience}.

Our key insight is that an object graph~\cite{Gyssens_ObjectGraphs},
i.e., in-memory program state available at the execution time, can be
seen as a property graph.  We believe that being able to query object
graphs during development, testing, and debugging will substantially
extend the power of programming languages and tools.


We present \emph{\underline{O}bject-\underline{G}raph
Pr\underline{o}gramming} (\Tool) that enables querying and updating an
object graph via declarative queries.
\Tool treats a given object graph as a graph database: each node in
the graph is an object (e.g., an instance of a class or an instance of
a meta class) and each edge is a relation between objects (e.g., a
field of one object references another object).
We leverage \Cypher
as \Tool's query language.
This gives rise to endless opportunities to leverage \Tool for programming,
analyses, and tool development.
We describe several potential use cases where \Tool can be applied.

\Tool provides a powerful and expressive way for writing assertions
and program invariants~\cite{Pengyu_Deuterium, MilicevicETAL11Squander}.  Assertions
written using \Tool not only can access the local program state, but
they can also check any aspect of the dynamic state of a program.

For example, we could assert that there is never more than one
instance of a specific (singleton) class.
Moreover, like \LINQ, \Tool can be used for querying collections of
data and even implementing these collections.
Unlike \LINQ, \Tool, at the moment, requires developers to cast their
results to appropriate type, as we do not guarantee type safety, which
is similar to working with the \Code{java.sql}~\cite{Java_SQL} package.
At the same time, \Tool can query any collection (e.g., an n
dimensional array) without requiring a user to implement any specific
interface.
 
In this paper, we show the power of \Tool with these two use cases,
but we envision many further benefits and potential uses.
 
For example, \Tool could facilitate dynamic program analyses. For instance, a
common task for tools that detect flaky
tests~\cite{Darko_TACAS21,Parry_FlakyTests,Bell_FlakyTests} (due to
test order dependency) is to check that the program state is the same
at the beginning of each test.

A common subtask is to find all objects reachable from static
variables.  Using \Tool, we can write a query to get all reachable
objects as \Code{query(``MATCH\;(n\;\{[]1\})-[*]\rightarrow(m)\;RETURN\;m",\;roots)}
starting from roots.
For another example, \Tool could be used to introspect the system, such as finding all
objects of a given class that satisfy a desired set of properties.
Unlike reflection, which is frequently used to discover relations
among objects and meta classes via imperative traversal of object
graphs, e.g., serialization
code~\cite{Java_Serialization_Specification, Braux_Serialization,
  Breg_Serialization_XStream, Philippsen_Serialization_XStream,
  Philippsen_Serialization_XStream2}, \Tool can help find these
relations via queries over instances in memory and instances of meta
classes (assuming they are available as part of the object graph like
in Java).
 
We discuss these and further benefits in Section~\ref{sec:limitations}.


We prototyped \Tool for the Java programming language in two ways:
(a)~by translating an object graph into a Neo4j database on which we
run \Cypher queries (\ToolNeo), and (b)~by implementing our own
in-memory graph query engine that directly queries the object graph
(\ToolMemory).
The former enables us to harvest the full power of a mature graph
database, including the highly optimized query engine and
visualization capabilities.
However, the translation cost from object graph can be high even with
a number of optimizations that we developed, and this approach
requires extra memory (disk).
Since, at the moment, we do not leverage stored graph databases for
any offline analysis, we developed a second prototype that works on
the object graph in memory.
This approach requires close to zero extra memory, but comes with
significant engineering effort.


We evaluate the applicability and robustness of \Tool by rewriting
\NumAssertions assertions from \NumProjects popular open-source
projects available on GitHub. Furthermore, we implemented a
number of methods from several classes using \Tool.
We report execution time for both prototypes.  Our results highlight
substantial performance benefits of in-memory implementation.


\noindent
The key contributions of this paper include:

\begin{itemize}[topsep=3pt,itemsep=4pt,partopsep=3pt,parsep=0ex,leftmargin=*]
\item[$\star$] \textbf{Idea}.
  \Tool introduces a new view of the runtime state of a program and
  provides a novel way by which such a state can be queried and
  modified.  \Tool offers developers a blend of imperative and
  declarative programming abstractions to manipulate the program
  state, increasing the expressivity of a programming language which
  implements \Tool's paradigm.  Although \Tool can be used to replace
  many statements (even a single field access), it is best suited for
  tasks that include traversal of objects and metadata, such as
  introspecting system state, writing assertions and
  invariants, and implementing linked data structures.
\item[$\star$] \textbf{Formalization}.
  We formalize \Tool by giving a small-step operational semantics to
  Featherweight Java~\cite{Igarashi_Featherweight} in terms of
  property graphs. This formalization captures the core of our
  translation to \Neo and can form the foundation for
  future projects that require reasoning about correctness, such as
  query optimizations.
\item[$\star$] \textbf{Implementation}.
  We implement two prototypes of \Tool by (a)~translating Java's
  object graph to an off-the-shelf graph database, and (b)~by
  implementing from-scratch-in-memory graph query engine that directly
  queries the object graph.
  Although our focus was
  on features supported by \Tool and not on its performance, we
  describe several optimizations for both translation and in-memory
  traversal.
\item[$\star$] \textbf{Evaluation}.
  We evaluated the robustness of our prototypes and compared their
  performance by rewriting a large number of assertions that are
  already available in popular open-source projects.  Focusing on
  assertions simplified the selection of target statements for the
  evaluation and enabled us to scale our experiments.
  We also implemented a number of methods from several
  classes in popular open-source projects.
\end{itemize}

\noindent
\Tool is publicly available at: \\
\url{https://github.com/EngineeringSoftware/ogo}.

\section{Examples}%
\label{sec:example}%

We demonstrate the expressive power of declarative queries for
analyzing program state by using two examples, such that each example
illustrates a different aspect of the framework: (1) creation of
instances, relations between instances and object graph pattern
matching;
(2) implementing instance methods (\Code{containsKey}) of \Java
Collections framework class (\Code{java.util.HashMap}).

\subsection{Creating and Querying Object Graphs}

\begin{figure*}[t]%
	\subcaptionbox{%
		\FigCreateDSInstantiateCaption%
		\label{figure:create_ds_example_instantiate}}{%
		\hspace*{-8mm}%
		\input{figs/example/create-ds/fig-create-ds-create-instance-and-relation}%
	}%
	\hspace*{1cm}%
	\captionsetup[subfloat]{%
		width=0.24\textwidth,%
		aboveskip=2pt,%
		belowskip=4pt,%
		justification=raggedright%
	}%
	\subcaptionbox{%
		\FigCreateDSMatchGenericReachabilityCaption%
		\label{figure:create_ds_example_match_generic_reachability}}{%
		\hspace*{-1cm}%
		\definecolor{bTreeColor}{RGB}{49, 168, 247}
\definecolor{matchColor}{RGB}{247, 47, 58}
\tikzstyle{instanceNode}=[circle,fill=black!91,thick,minimum size=3.5mm, text=white, text width=1.5mm, inner sep=0pt]
\tikzstyle{matchedNode}=[circle,fill=matchColor!77,thick,minimum size=3.5mm, text=white, text width=1.8mm, inner sep=0pt]
\tikzstyle{class}=[circle,fill=black!47,thick,minimum size=3.5mm]
\tikzstyle{instanceofEdge}=[-stealth,black!47]
\tikzstyle{instanceBinaryTree}=[circle,fill=bTreeColor!91,thick,minimum size=3.5mm]
\tikzstyle{codeBlock}=[rectangle, minimum width=0.6cm,minimum height = 1cm]
\newsavebox\ogoCodeBlockGenericMatch
\begin{lrbox}{\ogoCodeBlockGenericMatch}
	\lstinputlisting[language=small-java-pretty]{code/code-create-ds-bTree-generic-match.java}
\end{lrbox}
\begin{tikzpicture}[bend angle=45, font=\scriptsize]%
  \draw[dashed] (-0.5,1.5) -- (-0.5,4.5);
  \draw[dashed] (4.5,1.5) -- (4.5,4.5);
	\node[matchedNode] (classInst) at (2,4) {m};%
	\node[matchedNode] (classInstBTree) at ($ (classInst) + (10:0.7) $) {m};%
	\node[instanceNode,label={[yshift=0.3em]below:4}] (root) at ($ (classInst) + (290:1) $) {n}%
	edge [instanceofEdge] (classInst);%
	\node[matchedNode,label={[yshift=0.3em]below:5}] (rightChild) at ($ (root) + (335:1) $) {m}%
	edge [instanceofEdge] (classInst)%
	edge [stealth-] node[below, sloped] {right} (root);%
	\node[matchedNode,label={[yshift=0.3em]below:2}] (leftChild) at ($ (root) + (215:1.5) $) {m}%
	edge [instanceofEdge] (classInst)%
	edge [stealth-] node[below, sloped] {left} (root);%
	\node[matchedNode,label={[yshift=-0.3em]above:1}] (leftLeftChild) at ($ (leftChild) + (130:1.5) $) {m}%
	edge [instanceofEdge] (classInst)%
	edge [stealth-] node[below, sloped] {left} (leftChild);%
	\node[matchedNode,label={[yshift=-0.3em]above:3}] (leftRightChild) at ($ (leftChild) + (110:1) $) {m}%
	edge [instanceofEdge] (classInst)%
	edge [stealth-] node[above, sloped] {right} (leftChild);%
	\node[matchedNode] (bTree) at ($ (classInst) + (350:1.3) $) {m}%
	edge [-stealth] node[above, sloped] {root} (root)%
	edge [instanceofEdge] (classInstBTree);%
	\node[codeBlock] (cypherCode) at ($ (classInst) + (270:4) $)  {\usebox\ogoCodeBlockGenericMatch};%
\end{tikzpicture}%
%
	}%
	\hspace*{0.5cm}%
	\subcaptionbox{%
		\FigCreateDSMatchSpecificReachabilityCaption%
		\label{figure:create_ds_example_match_specific_reachability}}{%
		\hspace*{-10mm}%
		\definecolor{bTreeColor}{RGB}{49, 168, 247}
\definecolor{matchColor}{RGB}{247, 47, 58}
\tikzstyle{instanceNode}=[circle,fill=black!91,thick,minimum size=3.5mm, text=white, text width=1.5mm, inner sep=0pt]
\tikzstyle{matchedNode}=[circle,fill=matchColor!77,thick,minimum size=3.5mm, text=white, text width=1.8mm, inner sep=0pt]
\tikzstyle{class}=[circle,fill=black!47,thick,minimum size=3.5mm]
\tikzstyle{instanceofEdge}=[-stealth,black!47]
\tikzstyle{instanceBTree}=[circle,fill=bTreeColor!91,thick,minimum size=3.5mm]
\tikzstyle{codeBlock}=[rectangle, minimum width=0.6cm,minimum height = 1cm]

\newsavebox\ogoCodeBlockSpecificMatch
\begin{lrbox}{\ogoCodeBlockSpecificMatch}
	\lstinputlisting[language=small-java-pretty]{code/code-create-ds-bTree-specific-match.java}
\end{lrbox}

\begin{tikzpicture}[bend angle=45, font=\scriptsize]
	\node[class] (classInst) at (2,4) {};
	\node[class] (classInstBTree) at ($ (classInst) + (10:0.7) $) {};
	\node[instanceNode,label={[yshift=0.3em]below:4}] (root) at ($ (classInst) + (290:1) $) {n}
	edge [instanceofEdge] (classInst);
	\node[instanceNode,label={[yshift=0.3em]below:5}] (rightChild) at ($ (root) + (335:1) $) {}
	edge [instanceofEdge] (classInst)
	edge [stealth-] node[below, sloped] {right} (root);
	\node[instanceNode,label={[yshift=0.3em]below:2}] (leftChild) at ($ (root) + (215:1.5) $) {}
	edge [instanceofEdge] (classInst)
	edge [stealth-] node[below, sloped] {left} (root);
	\node[matchedNode,label={[yshift=-0.3em]above:1}] (leftLeftChild) at ($ (leftChild) + (130:1.5) $) {m}
	edge [instanceofEdge] (classInst)
	edge [stealth-] node[below, sloped] {left} (leftChild);
	\node[instanceNode,label={[yshift=-0.3em]above:3}] (leftRightChild) at ($ (leftChild) + (110:1) $) {}
	edge [instanceofEdge] (classInst)
	edge [stealth-] node[above, sloped] {right} (leftChild);
	\node[instanceBTree] (bTree) at ($ (classInst) + (350:1.3) $) {}
	edge [-stealth] node[above, sloped] {root} (root)
	edge [instanceofEdge] (classInstBTree);
	\node[codeBlock] (cypherCode) at ($ (classInst) + (270:3.85) $)  {\usebox\ogoCodeBlockSpecificMatch};
\end{tikzpicture}%
	}%
    \vspace{-5mm}%
	\definecolor{bTreeColor}{RGB}{49, 168, 247}
\definecolor{matchColor}{RGB}{247, 47, 58}
\tikzstyle{instanceNode}=[circle,fill=black!91,thick]
\tikzstyle{instanceRel}=[circle,fill=black!0]
\tikzstyle{instanceBTree}=[circle,fill=bTreeColor!91,thick]
\tikzstyle{class}=[circle,fill=black!47,thick]
\tikzstyle{instanceofEdge}=[-stealth,black!47]
\tikzstyle{matchedNode}=[circle,fill=matchColor!77,thick]
\begin{tikzpicture}[bend angle=45, font=\scriptsize, overlay]
	\node [class,label=below:java.lang.Class instance] (classDescription) at (-6.2,6.5) {};
	\node [instanceNode,label=below:Node instance] (instanceNodeDescription) at ($ (classDescription) + (0:3) $){};
	\node [instanceBTree,label=below:BinaryTree instance] (instanceBTreeDescription) at ($ (instanceNodeDescription) + (0:3) $){};
	\node [matchedNode,label=below:Matched instance] (matchedNodeDescription) at ($ (instanceBTreeDescription) + (0:3) $){};
	\node [class] (classInst) at ($ (matchedNodeDescription) + (0:3) $){};
	\node [instanceRel, label={[xshift=-1.5em]below:instanceof Relation}] (inst) at ($ (classInst) + (0:1) $){}
	edge [instanceofEdge] (classInst);
\end{tikzpicture}%
	\caption{\FigCreateDSCaption\label{figure:create_ds_example}}
        \vspace{-10pt}
\end{figure*}%

The binary tree is a rooted ordered tree with each of its nodes having
at most 2 children. We demonstrate the versatility of \Tool by
leveraging declarative queries to construct a binary tree and query it
for complex patterns. We also use this example to introduce the syntax
of the \Cypher query language~\cite{Francis_Cypher}.

A \Java implementation of the binary tree is given in
Figure~\ref{figure:korat_example_btree_def}. An instance of
\Code{BinaryTree} contains a reference field \Code{root} of type
\Code{Node} (short for \Code{BinaryTree\$Node}), the root node of the
tree and a primitive \Code{int} field \Code{size} that tracks the
total number of nodes in the tree. An instance of \Code{Node} contains
references to its left and right child nodes also of type \Code{Node}
and stores an integer value in its primitive \Code{int} field
\Code{value}.

Constructing a binary tree using \Tool is given in
Figure~\ref{figure:create_ds_example_instantiate}. We use the
\Code{queryObject} method of \Tool to execute the given \Cypher query
string.
Based on the \Cypher grammar, the query contains six clauses. The
first two are \Code{CREATE} clauses (write to database/object graph),
the next three are \Code{MERGE} clauses (write or read from
database/object graph) and finally, a \Code{RETURN} clause (defines
expressions to be returned).

The first \Code{CREATE} clause
(lines~\ref{line:create_btree_node_instances_a_begin}-\ref{line:create_btree_node_instances_a_end})
creates 5 \Code{Node} instances.
These instances are assigned variable (which is a term used
  in the \Cypher syntax) names \textbf{a-e} for referencing in
followup clauses. The expansion of the positional arguments \Code{@1,
  @2} is described in Table~\ref{fig:positional:arguments} and are
replaced with the fully qualified class name (\Code{BinaryTree\$Node}
and \Code{BinaryTree} respectively) of the arguments following the
query string.
In \Cypher syntax, these are termed \Code{LABELS}. \Java 
non-primitive types are mapped to string \Code{LABELS} of nodes 
in our graph model.
Consequently, the \Code{CREATE} clause creates instances of the
specified type. The \Code{LABELS} are followed by node properties
(\Code{\{<prop\_name>:<prop\_value>, ...\}}) which are key-value pairs
that map to the represented object's primitive/String fields and their 
values. In \Code{CREATE}, they assign the fields of the created 
instance to the specified values. Thus, the \Code{value} field of 
instances \textbf{a-e} is assigned with values 1-5 respectively.
The second \Code{CREATE}
(line~\ref{line:create_btree_btree_instance_a}) creates an instance of
\Code{BinaryTree} (assigned \textbf{f}, this can be subsumed into the
first \Code{CREATE} but is divided for clarity).

The \Code{MERGE} clauses
(line~\ref{line:merge_btree_node_relations_a}) are used to create
relationships among the \textbf{a-e}. Since the binary tree is a
directed graph, we create directed relationships. \Code{LABELS} can
also be specified for relationships. In our graph model of the heap,
the relationship between a referrer and a referee instance is labelled
by the reference field corresponding to the referee instance in the
referrer instance's class. We use the reference fields \Code{left} and
\Code{right} as labels for the relationships between \textbf{a-e}. For
e.g., \Code{(b)\leftarrow[:left]-(c)-[:right]\rightarrow(d)}
translates to assigning \textbf{b} and \textbf{d} as the left and
right child of \textbf{c}. The \Code{RETURN} clause returns a
reference to the \Code{BinaryTree} instance \textbf{f} to prevent the
objects from being garbage collected.

We next describe querying the binary tree by discussing two patterns.
The first pattern investigates general reachability of objects from a
given object. Such queries are of importance to the problem domains of
Aliasing, Confinement and Ownership. For instance, if a node in the
binary tree was owned by another instance outside the confinement of
the binary tree instance then the aliased node could potentially be
mutated, leading to undesirable
outcomes. Figure~\ref{figure:create_ds_example_match_generic_reachability}
shows a query which returns all objects reachable from the \Code{root}
\Code{Node} instance. The query contains a \Code{MATCH}
(line~\ref{line:match_btree_generic_match}) and \Code{RETURN}
clauses. The \Code{MATCH} clause matches all paths in the heap's
object graph satisfying the given pattern \Code{(n
  \{\$1\})-[*]-(m)}. The positional argument \Code{\$1} expands to a
unique identifier belonging to the first argument following the query
and is used to uniquely identify an object on the heap.
The matched \Code{root} \Code{Node} instance is assigned variable \textbf{n}. The
pattern neither specifies a label nor a direction for the relationship
from \textbf{n} and hence all references from and to \textbf{n} are
considered. Furthermore, the \textbf{*} implies that the
referee/referrer instances can be any number of hops away from \textbf{n}. Thus,
this pattern matches the set of all objects reachable to and from
\textbf{n} under transitive closure and is assigned \textbf{m}.
This can be used to identify confinement issues. The pattern given in
Figure~\ref{figure:create_ds_example_match_specific_reachability}
describes a situation where we are interested in querying the binary
tree for existence of a node with a certain value (\Code{value}=1) and
a certain distance (2 hops) away from the root node. This is a more
constrained query than the former and the pattern
(lines~\ref{line:match_btree_specific_match_begin}-\ref{line:match_btree_specific_match_end})
is more specific. The relationship is now directed
(\Code{-[]\rightarrow}), labelled and with fixed distance so only
instances referenced by \textbf{n} and through reference fields
\Code{left} or \Code{right} that are 2 (\Code{[*2]}) hops away from
\textbf{n} are considered.

\subsection{Implementing \Java Library Methods}

\begin{figure}[t]%
  \subcaptionbox{\FigJavaDSDefCaption \label{figure:java_ds_example_def}}{%
\begin{lstlisting}[language=small-java-pretty]
public class HashMap<K,V> extends AbstractMap<K,V>
	                     implements Map<K,V>,Cloneable,Serializable {
    transient Node<K,V>[] table;
    static class Node<K,V> implements Map.Entry<K,V> {
        final int hash; final K key; V value; Node<K,V> next;
    }
}
\end{lstlisting}%
  }%
  
  \subcaptionbox{\FigJavaDSImperativeCaption \label{figure:java_ds_example_imperative}} {%
    \hspace*{-3mm}%
\begin{lstlisting}[language=small-java-pretty]
public boolean containsKey(Object key) {
    return getNode(hash(key), key) != null;
}
final Node<K,V> getNode(int hash, Object key) {
    Node<K,V>[] tab; Node<K,V> first, e; int n; K k;
    if ((tab = table) != null && (n = tab.length) > 0 &&
            (first = tab[(n - 1) & hash]) != null) {
        if (first.hash == hash && // always check first node
            ((k = first.key) == key || (key != null && key.equals(k))))
            return first;
        if ((e = first.next) != null) {
            if (first instanceof TreeNode)
                return ((TreeNode<K,V>)first).getTreeNode(hash, key);
            do {
                if (e.hash == hash && ((k = e.key) == key ||
                     (key != null && key.equals(k)))) return e;
            } while ((e = e.next) != null);}
	} return null;}
\end{lstlisting}%
  }%
  
  \subcaptionbox{\FigJavaDSOGOCaption \label{figure:java_ds_example_ogo}} {%
    \hspace*{-9mm}%
\begin{lstlisting}[language=small-java-pretty]
public boolean containsKey(Object key) {
    return queryBool(
       "MATCH ({$1})-[:table]->()-[*]->()-[:key]->(n) MATCH (m {$2})!<\label{line:hashmap_match_clause}>! 
        WHERE n.`equals`(m) = !<\textcolor{britishracinggreen}{true}>! RETURN COUNT(n) <> 0", this, key);
}
\end{lstlisting}%
  }%
  \vspace{-10pt}%
  \caption{\FigJavaDSCaption\label{figure:java_ds_example}}%
  \vspace{-13pt}%
\end{figure}%

The \Java collections framework provides a rich collection of data
structures supported natively by the \Java platform.
We show how \Tool can be used to manipulate these objects by
considering the example of implementing methods available
  in the \Java collections' class
\Code{java.util.HashMap}. The \Code{HashMap} stores data as key-value
pairs where every stored value is mapped to a unique key.

A snippet of the \Code{HashMap} class is given in
Figure~\ref{figure:java_ds_example_def}. The reference field
\Code{table} contains all the entries in the map. The method
\Code{containsKey} of the \Code{HashMap} class checks if a given key
is present in the map. A purely imperative implementation of the
\Code{containsKey} method is given in
Figure~\ref{figure:java_ds_example_imperative}. \Tool implementation
is shown in Figure~\ref{figure:java_ds_example_ogo}. The \Cypher query
used contains 2 \Code{MATCH} clauses
(line~\ref{line:hashmap_match_clause}), the first clause matches all
instances reachable from \Code{table} that correspond to the reference
field \Code{key} (defined in the static nested class \Code{Node} in
\Code{HashMap}) and refers to these set of instances as \textbf{n},
the second clause matches the instance passed in as an argument to the
\Code{containsKey} method. The \Code{WHERE} clause is used to filter
the set \textbf{n} based on the result of the \Code{equals} method
(overriden or inherited from \Code{java.lang.Object}). If the
\Code{equals} method evaluates to \Code{true} for at
  least one instance in \textbf{n} and for \textbf{m} then
the cardinality of set \textbf{n} is non-zero after this clause is
completed and hence the \Code{RETURN} clause returns \Code{true}.

The examples show a glimpse of the potential of \Tool: it provides
access to any object in memory, at any point regardless of access
specifiers, through a declarative API.
Similar to this example, one could envision numerous other potential
applications of \Tool, a point which we return to in
Section~\ref{sec:limitations}.

\section{Framework}
\label{sec:framework}

We first present \Tool's API design (Section~\ref{sec:api}) followed
by a high-level overview of the workings of \Tool
(Section~\ref{sec:overview}).
Next, we formally describe mapping of the object graph in 
the JVM heap memory to a property graph, as well as translation to
\Neo (Section~\ref{sec:formal}).
Lastly, we conclude the section with a description of our
implementation and optimizations details
(sections~\ref{sec:implementation} and \ref{sec:optimizations}).

\subsection{API}
\label{sec:api}

\begin{figure} [t]%
  \makebox[.45\textwidth]{%
\begin{lstlisting}[language=small-java-pretty]
public static ResultSet query(Object root, String fmt, Object... values); !<\label{line:bounded_query}>!
public static ResultSet query(String fmt, Object... values); !<\label{line:unbounded_query}>!
public static Object queryObject(Object root, String fmt, Object... values);!<\label{line:bounded_query_object}>!
public static Object queryObject(String fmt, Object... values);
public static boolean queryBool(Object root, String fmt, Object... values); !<\label{line:other_query_begin}>!
public static boolean queryBool(String fmt, Object... values);
public static long queryLong(Object root, String fmt, Object... values); !<\label{line:bounded_query_long}>!
public static long queryLong(String fmt, Object... values); !<\label{line:other_query_end}>!
// similar for other primitive types
\end{lstlisting}}%
  \vspace{-3mm}%
  \caption{\FigOgoApiCaption}%
  \label{figure:api}%
  \vspace{-5mm}%
\end{figure}%

We begin by describing queries---their type, arguments, and return values---followed by a discussion of our API design
choices.

\MyPara{Queries}
We show (the most important parts of) \Tool's \API in
Figure~\ref{figure:api}.
The design choice of keeping the \API minimal is intentional, similar
to that available for working with relational databases such as
\Code{java.sql}. This allows developers acquainted with both Java
and \Cypher to be able to use \Tool with ease.

The highlight of the \API are two variadic \Code{query} methods
(lines~\ref{line:bounded_query} and \ref{line:unbounded_query} in
Figure~\ref{figure:api}), which we call \emph{\boundedq} and
\emph{\unboundedq}, respectively.  We describe each one in turn.

In case of a \emph{\boundedq}, the first argument of the method takes
an object (\Code{root}) that constraints query execution only to
objects reachable (under transitive closure of reference fields) from this \Code{root}.
Having the ability to specify only a subgraph of the entire object
heap enables two things: (a)~making localized queries, e.g., like we
did in
Figure~\ref{figure:create_ds_example_match_specific_reachability}, and
(b)~improve performance of \Tool, as we focus on traversal of only a
part of the entire object heap.
A nice side effect of \Code{root} having an \Code{Object} type is
that one can pass a collection that contains multiple roots in a
single invocation.
For example, if the goal is to find all objects reachable from several
static fields, one can add values
of all those static fields into a single collection and pass that
collection as the first argument to the \boundedq.

In case of an \emph{\unboundedq}, the query is executed on the entire
available object heap.
We had an example of such a query in Figure~\ref{figure:java_ds_example_ogo}.
(In that example an \unboundedq was used since the instances being matched
may not be related.)
Semantics for an \unboundedq are relatively less precise than for a 
\boundedq due to the dynamic nature of the \JVM, and the user has to accept
these unknowns if they use \unboundedqs.
Specifically, Java is a GC-enabled programming language and \Tool
\emph{does not} guarantee that objects returned by \unboundedqs will
not be garbage.
In the future, we might offer a more precise semantics for
\unboundedqs, e.g., ensuring that the result of a query reflects the
state as if GC completed its work (this will be trivial to offer during
a debugging session for example).
 Nevertheless, \unboundedqs can be valuable in countless examples,
e.g., checking if there are instances of classes that are not expected
to be instantiated or checking if at any point all instances of a
specific class have a field in a given range of values.

\begin{table}[t]%
	\TableFontSize%
	\caption{%
		\TablePositionalArgumentsCaption%
		\label{fig:positional:arguments}}%
	\begin{tabular}{lp{6.5cm}}%
		\toprule%
		\Code{\$} & Embed a unique id of an instance\\
		example     &%
\begin{lstlisting}[language=positional-argument-table]
query("MATCH (n {!<\color{britishracinggreen}\textbf{\$}>!1}) RETURN n.some_field", node)
\end{lstlisting}\\
		expansion   &%
\begin{lstlisting}[language=positional-argument-table]
query("MATCH (n {__id__:5777203}) RETURN n.some_field")
\end{lstlisting}\\
		\midrule
		\Code{@}  & Embed the fully qualified class name of an instance\\
		example     &%
\begin{lstlisting}[language=positional-argument-table]
query("MATCH (n: !<\color{britishracinggreen}\textbf{@}>!1) RETURN n", node)
\end{lstlisting}\\
		expansion   &%
\begin{lstlisting}[language=positional-argument-table]
query("MATCH (n:`com.!<\textcolor{britishracinggreen}{package}>!.Node`) RETURN n")
\end{lstlisting}\\
		\midrule
		\Code{[]} & Embed unique id of instances from an \texttt{Iterable} collection and return union of results \\
        example     &%
\begin{lstlisting}[language=positional-argument-table]
query("MATCH (n {!<\color{britishracinggreen}\textbf{[]}>!1}) RETURN n.some_field", nodes)
\end{lstlisting}\\
		expansion   &%
\begin{lstlisting}[language=positional-argument-table]
query("MATCH (n {__id__:8898223}) RETURN n.some_field")!<$\cup$>!
query("MATCH (n {__id__:5777203}) RETURN n.some_field")...
\end{lstlisting}\\
		\bottomrule%
	\end{tabular}%
	\vspace{-8pt}%
\end{table}%

\MyPara{Arguments}
Both query methods in our \API share the remaining arguments:
formatting string (\Code{fmt}) and values.
The formatting string in its simplest form is just a \Cypher query
such as \Code{"MATCH\, (n:\textasciigrave
  java.util.ArrayList\textasciigrave)\, RETURN\, n"}.
To enable constraining queries by embedding runtime values, we
introduce several kinds of positional arguments; the values are
provided from the third argument onward.  We show kinds of positional
arguments in Table~\ref{fig:positional:arguments}.  For each kind, we
show an example and the expansion once the formatting is complete.
We support embedding the unique id of an instance (\Code{\$}), fully
qualified class name of an instance (\Code{@}), or doing a union of
query results when we run a query on each element of a
(\Code{Iterable}) collection (\Code{[]}).

\MyPara{Return value}
The result of each query is an instance of
\Code{ResultSet}.  \Code{ResultSet}~\cite{Java_SQL} is available
in Java as an interface and a common structure for storing the results
of queries; similar structure is used in other programming languages.
Via the resulting instance one can access columns (e.g.,
\Code{getArray(int\, columnIndex)}), get the current row number
(e.g., \Code{getRow()}), etc.
Anybody already familiar with working with relational databases from
Java would be familiar with the structure.

For convenience, we introduced several query methods that return a
specific type (lines~\ref{line:other_query_begin}-\ref{line:other_query_end}).
The only difference is that these methods cast/extract the result as a
single value from the \Code{ResultSet}; many assertions or queries
that use \Code{COUNT} would end up benefiting from this shorter
version.  As for the naming, we followed similar convention as the
\Code{Unsafe} class~\cite{UnsafeClass}.

\MyPara{Design decisions}
Similar to working with SQL strings and \Code{java.sql}, \Tool does
not statically type check expressions.  Thus, a dynamic
\Code{CastClassException} might be raised if a wrong value is passed
to one of the query calls.
Alternatively, rather than specifying the \Cypher query as a string
explicitly, the \Cypher DSL~\cite{GerritCypherDSL2020} could be used
instead.  We leave this integration for future work.

Moreover, our \API is not designed to be thread safe.  Namely, the
developer is responsible to ensure that appropriate locks are held
when querying the (sub) object graph.  This approach offers more
flexibility without being different than implementing any code snippet
imperatively.

\begin{figure*}[t]
  \includegraphics[scale=0.88]{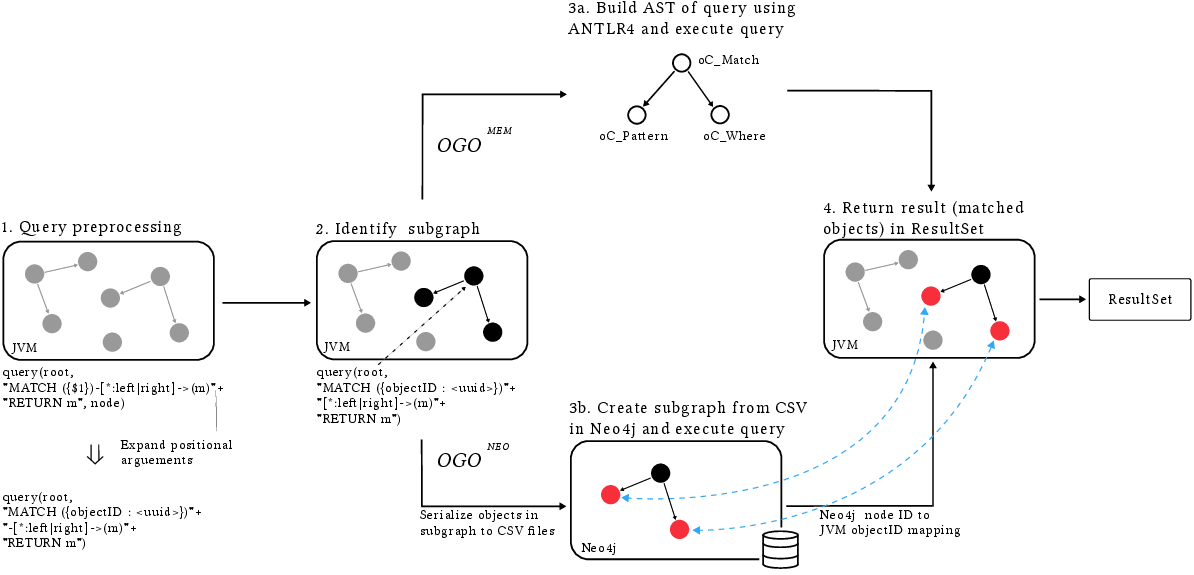}
  \caption{Overview of \Tool (and two implementations: \ToolNeo and \ToolMemory).\label{fig:overview}}
\end{figure*}

\subsection{Overview of \Tool Steps}
\label{sec:overview}

Figure~\ref{fig:overview} shows the high level overview of the working
of \Tool. The figure illustrates steps taken by both of our
implemented prototypes (\ToolNeo and \ToolMemory),
and highlights the differences between the two.

\Tool flow starts once a query method is invoked, as described in the
previous section. In addition to the query, \Tool implicitly takes as
input the entire state of the program. In the first step, \Tool
processes the formatting string and builds the actual query to be
executed.
This step is straightforward and includes simple string manipulations
and object id discoveries (if the user has any collection in the
formatting string, e.g., \Code{[]}, \Tool builds a batch query).
In the second step, \Tool uses a \JVMTI agent (and the root object if given)
to identify objects that are in the (sub)graph of interest. 
Note that a highly optimized system would not traverse the objects
before analyzing the given query.
Once the second step is done, the execution for \ToolNeo and
\ToolMemory diverge.

\ToolMemory, in the third step, builds intermediate representations of
the query (\AST among others) and executes the query as per the
\Cypher semantics~\cite{Francis_CypherSemantics}.  In the final step,
\ToolMemory collects the results into a newly allocated
\Code{ResultSet}, which is the final result of the query execution.

Unlike \ToolMemory, \ToolNeo, in the third step, \emph{translates}
(Section~\ref{sec:formal}) objects of interest and their relations
into a format accepted by \Neo for batch data loading.
In step four, \ToolNeo passes the query to a \Neo database running in
a separate \JVM and takes the result of the execution, which
are node IDs when the result type is non-primitive and primitive 
values when the result type is primitive.
In the final step, it processes these node IDs to build the final
result, which are the values known to the \JVM. Primitive values
remain as they are returned by \Neo. On the other hand, non-primitive
values are mapped to object IDs, which are then used to fetch objects
and build the final \Code{ResultSet} returned to the user. \ToolNeo
and the \Neo database are implemented as RMI client and server
applications respectively. This prevents polluting the \JVM heap with
irrelevant \Neo database objects.

\subsection{Translation}
\label{sec:formal}

In this section we describe our translation from the \JVM heap to a
\Neo graph database, by formalizing graph databases and presenting an
operational semantics for \Featherweight Java in terms of this
formalization.
\MyPara{Graph databases, formally}
A graph database is a {\em directed multigraph}:
a pair $\Nods \times \Rels$ of {\em
  nodes}, the main entities of the graph, and {\em relationships}, the
edges of the graph denoting directed connections between nodes.
A {\em node} is a pair $\Labs \times \Props$ of a {\em label}\footnote{Corresponds to the fully
qualified class name of the instance represented by the node.}, drawn
from some abstract domain that serves as the type of the node, and
{\em properties}, a map from string keys to string values.
A {\em relationship} corresponds to the edges of the multigraph: it
has a start and end node, a label,
and a key-value property map.

\begin{figure}[t]
  \subcaptionbox{%
    \FigKoratBTreeDefCaption%
    \hspace*{10cm}%
    \label{figure:korat_example_btree_def}}{%
    \hspace*{1mm}%
\begin{lstlisting}[language=small-java-pretty]
class BinaryTree {
   private Node root;
   private int size;
   static class Node {
      private Node left;
      private Node right;
      int value;
   }
}
\end{lstlisting}%
  }%
  \hfill%
  \subcaptionbox{%
    \FigTranslationCodeCaption%
    \label{figure:translation_test_class}}{%
    \hspace*{5mm}%
\begin{lstlisting}[language=small-java-pretty]
class BinaryTreeTest{
   void test(){
      BinaryTree b = new BinaryTree();
      Node l = new Node(4);
      b.root = new Node(l, null, 5);
      !<\textcolor{britishracinggreen}{\textbf{($\ast$ POINT $\ast$)}}\label{line:translation_point}>!
      query(...)
   }
}
\end{lstlisting}%
  }%
  \\  
  \subcaptionbox{%
    \FigTranslationFigCaption%
    \label{figure:translation_object_graph}}{%
    \hspace*{-5mm}%
    \definecolor{bTreeColor}{RGB}{49, 168, 247}
\tikzstyle{instanceNode}=[circle,fill=black!91,thick,minimum size=4mm, text=white, text width=1mm, inner sep=0pt]
\tikzstyle{class}=[circle,fill=black!47,thick,minimum size=4mm]
\tikzstyle{localVar}=[rectangle,black!50]
\tikzstyle{instanceofEdge}=[-stealth,black!47]
\tikzstyle{localVarEdge}=[-stealth,dashed,black!47]
\tikzstyle{instanceBinaryTree}=[circle,fill=bTreeColor!91,thick,minimum size=4mm, text=white, text width=1mm, inner sep=0pt]
\begin{tikzpicture}[bend angle=45, font=\footnotesize]
	\node[class,label={[yshift=0.3em]below:\{\;...\;\}}] (classInst2) at (1,4) {};
	\node[class,label={[yshift=0.3em]below:\{\;...\;\}}] (classInst2BTree) at ($ (classInst2) + (0:2.5) $) {};
	\node[instanceNode,label={[yshift=0.3em]below:\{\;value:5\;\}}] (root2) at ($ (classInst2) + (315:2.4) $) {}
	edge [instanceofEdge] node[above, sloped] {instanceof} (classInst2);
	\node[instanceNode,label={[yshift=0.3em]below:\{\;value:4\;\}}] (leftChild2) at ($ (root2) + (180:3.5) $) {}
	edge [instanceofEdge] node[above, sloped] {instanceof} (classInst2)
	edge [stealth-] node[above, sloped] {left} (root2);
	\node[instanceBinaryTree,label={[yshift=0.3em]below:\{\;size:2\;\}}] (bTree) at ($ (root2) + (0:3.5) $) {}
	edge [-stealth] node[above, sloped] {root} (root2)
	edge [instanceofEdge] node[above, sloped] {instanceof} (classInst2BTree);
    \node[localVar,label={[yshift=-0.3em]above:local var}] (lvar1) at ($ (bTree) + (90:1.7) $) {}
	edge [localVarEdge] node[right] {b} (bTree);
    \node[localVar,label={[yshift=-0.3em]above:local var}] (lvar2) at ($ (leftChild2) + (90:1.7) $) {}
	edge [localVarEdge] node[left] {l} (leftChild2);
\end{tikzpicture}%
  }%
\vspace{-3mm}%
\caption{\FigTranslationCaption}%
\label{figure:translation}%
\vspace{-5mm}%
\end{figure}%

\MyPara{The \Java heap as a graph database, by example}
Following the binary tree example presented in
Section~\ref{sec:example}, consider a simple \Code{BinaryTree}
instance in Figure~\ref{figure:translation_test_class}, which defines
a \Code{Node l} with a \Code{value} field 4, and a
\Code{BinaryTree b} whose root is a \Code{Node} with \Code{l} as
its left subtree, no right subtree and a \Code{value} field of 5. A
pictorial representation of the property graph at the marked
\Code{POINT} is shown in
Figure~\ref{figure:translation_object_graph}.

\begin{itemize}[topsep=3pt,itemsep=3pt,partopsep=0ex,parsep=0ex,leftmargin=*]
\item {\bf Object instances.}
  Every object instance that has been
  allocated corresponds to a node in the property graph. It's label is
  the name of the object's class, and it's property set contains the
  values of any primitive fields or strings. In our example, the node
  corresponding to \Code{l} has the label \Code{Node} and a
  singleton property set that maps \Code{value} to \Code{4}, while
  the node corresponding to \Code{b} has the label \Code{BinaryTree}
  and a singleton property set that maps \Code{size} to value \Code{2}.
  
\item {\bf Fields.}
  Each reference field of an object corresponds to a
  relationship whose label is the name of the field, its origin is the
  node of the graph that corresponds to the object it belongs to, its
  destination is the object corresponds to the field's value, and its
  property set is empty. The \Code{root} field of \Code{b} is then
  an edge in the graph going from \Code{b} to the nameless node
  corresponding to the second allocation, whose \Code{left} field in
  turn points to \Code{l}.

\item {\bf Local variables.}
  Each local variable we introduce (such as
  \Code{n} and \Code{l}), gives rise to both a node in the graph
  whose label is $\local$ and whose property set is empty, and to a
  relationship whose label is the name of the variable from that node
  to the object corresponding to the local variable's
  value. 

\item {\bf Class information.}
  Each object is also related to a static definition of its class via
  an \Code{instanceof} relationship.  This allows us to capture, for
  example, static fields belonging to a class as its property set.
\end{itemize}

\begin{figure}[t]
\vspace{-3mm}%
\[
\begin{array}{l}
  \tL ::= \classd{\tC}{\tC}\\
  \tK ::= \constrd\\
  \tM ::= \methodd\\
  \tc ::= \tx = \tnew ~ \tC(\txs) | ~ \tx.\tf = \tx ~ | ~ \tx.\tm(\txs)\\
  \te ::= \tc; ~ \te ~ | ~ \treturn ~ \tx 
\end{array}  
\]
\vspace{-4mm}%
\caption{Syntax of Featherweight Java.}
\vspace{-4mm}%
\label{fig:Feather-syntax}
\end{figure}

\MyPara{Featherweight Java}
Featherweight Java, as introduced by \citet{Igarashi_Featherweight},
constitutes the object-oriented core of Java. Its syntax is shown in
Figure~\ref{fig:Feather-syntax}; it consists of: class declarations
such as $\classd{\tC}{\tD}$, which introduce a class $\tC$ with $\tD$ as its
superclass, $\tfs$ as its fields, $\tK$ as its constructor, and $\tMs$
as its methods; constructor declarations $\constrd$ which initialize
the fields of an instance of $\tC$; and method declarations $\methodd$
which define a method $\tm$ with arguments $\txs$ of types $\tCs$
whose body $\te$ returns a type $\tC$.
This setup allowed \citeauthor{Igarashi_Featherweight} to resolve
field ($\ifs$), method type ($\imt$), and method body ($\imb$) lookups
from a fixed class table in a straightforward manner, which we will
assume in the rest of the presentation.

Unlike Featherweight Java, where the bodies $\te$ are a single
$\treturn$ expression, we consider method bodies in assignment normal
form (ANF), as our aim is to formalize the small-step impact of
expressions as a graph rather than their big-step reductions and their
interactions with (sub)typing. Method bodies are therefore either
$\treturn$ expressions that return a variable, or sequences of
commands $\tc$ that either allocate a new object, assign to a field,
or invoke a method.

\begin{figure}
\begin{mathpar}
  \inferrule{
  }{
    \stepM{G}{\treturn ~ \tx}{G}
  } \quad
  \inferrule{
    \step{G}{\tc}{G'} \quad \stepM{G'}{\te}{G''} 
  }{
    \stepM{G}{\tc; \te}{G''}
  }
\end{mathpar}
\vspace{-5mm}%
\caption{Big-step reduction for Featherweight Java expressions.}
\label{fig:Feather-bigstep}
\vspace{-5mm}%
\end{figure}
\begin{figure}
  \begin{mathpar}
  \inferrule{
    (\Nods, \Rels) = G\\
    n_{\tx} = \Rels(\tx).3 \\
    n_{\ty} = \Rels(\ty).3 \\
    r = (\tf, n_{\tx}, n_{\ty}, \emptyset) \\
    G' = (\Nods, r \cup \Rels / \Rels(\tf, n_{\tx}, \cdot))\\
  }{
    \step{G}{\tx.\tf = \ty}{G'}
  }
  \\
  \inferrule{
    \imb(\tm) = \tys.e \\
    \stepM{G}{\te[\txs/\tys, \tx/\tthis]}{G'}
  }{
    \step{G}{\tx.\tm(\txs)}{G'}
  }\\
  \inferrule{
    n_{\tx} = (\local, \emptyset) \\
    n_{\tC} = (\tC, \emptyset) \\
    F = \fs (n_{\tC}, \tC(\txs)) \\ 
  }{
    \step{(\Nods, \Rels)}{\tx = \tnew ~ \tC(\txs)}
         {(\Nods \cup \{n_{\tx}, n_{\tC}\}, \Rels \cup F)}
  }
\end{mathpar}
\caption{Small-step reduction for Featherweight Java commands.}
\label{fig:Feather-smallstep}
\vspace{-5mm}%
\end{figure}
\begin{figure}
  \begin{mathpar}
    \inferrule{  
    \ifs(\tC) = \tfs \\
    \tC(\tys) = \tthis.\tfs=\tys \\
  }{
    \fs (n_{\tC}, C(\txs)) = \cup \{(f_i, n_{\tC}, \Nods(\tx_i).3), \emptyset\}
  }\\ 
  \inferrule{
    \tC ~ \textends ~ \tD \\
    \ifs(\tC) = \tfs \\
    \tC(\tys) = \tsuper(\tys_1^k); \tthis.\tfs=\tys_k \\
    F = \fs (n_{\tC}, D(\txs_1^k)) 
  }{
    \fs (n_{\tC}, C(\txs)) = F \cup \{(f_i, n_{\tC}, \Nods(\tx_{i+k}).3), \emptyset\}  
  }\\ 
  \end{mathpar}
  \vspace{-1cm}%
\caption{Calculation of fields relationships.}
\label{fig:Feather-fields}
\vspace{-10pt}%
\end{figure}

What we are ultimately interested in is modelling the shape of the
object graph after each command. To that end, we introduce two
(mutually recursive) judgements: a small-step judgement that relates a
command $\tc$ and an input graph $G$ to the resulting graph $G'$
and a big-step judgement that relates an entire expression $\te$ and an
input graph $G$ to the resulting graph $G'$:
$$\step{G}{\tc}{G'} \quad \quad \quad \stepM{G}{\te}{G'}.$$
The latter is unsurprising (Figure~\ref{fig:Feather-bigstep}), as
expressions are either returns (with no effect on the graph) or
sequences of commands (which compose the effects of the individual
commands).
The former is much more intricate, as it actually involves
manipulating the object graph via adding nodes, relationships, and
labels (Figure~\ref{fig:Feather-smallstep}).

To assign a variable $y$ to the $f$ field of a variable $x$ in a graph
$G = (\Nods, \Rels)$, we first lookup the relationships $\Rels(y)$ and
$\Rels(x)$, whose labels are $y$ and $x$ respectively, and extract the
nodes $n_{\ty}$ and $n_{\tx}$ corresponding to the objects these
fields currently point to in the object graph (the third component of
the relationship tuple). We then create a new relationship $r$ whose
label is $f$, pointing from $n_{\tx}$ to $n_{\ty}$ with an empty
object map. Finally, we update the object graph to include this new
relationship while removing the previous relationship corresponding to
the $f$ field of $\tx$.

To invoke a method $m$ of a variable $x$ with some arguments $\txs$,
we first lookup the body of the method (using $\imb$ as defined in
\citeauthor{Igarashi_Featherweight}), which is an expression $e$
parameterized by arguments $\tys$. We then substitute $\txs$ for
$\tys$ and $x$ for $\tthis$ in $e$ and use the big-step judgment for
expressions to construct the resulting object graph $G'$.

Finally, to create a new object $C$ by invoking its constructor with
some arguments $\txs$ and asssigning this new object to a fresh local
variable $\tx$, we need to extend the graph with two new objects, one
corresponding to $\tx$ (whose label is $\local$ and whose property map
is empty), and one corresponding to the newly created object (whose
label is $C$ and whose property map is empty). Then we need to create
a collection of field relationships
(Figure~\ref{fig:Feather-fields}) to account for the initialization of
the new object. We do that via a helper meta-function $\fs$,
identifying two cases:
\begin{itemize}[topsep=3pt,itemsep=3pt,partopsep=0ex,parsep=0ex,leftmargin=*]
\item If $C$ does not extend another class, then its constructor does
  not involve a call to $\tsuper$ and is just a sequence of field
  initializations $\tthis.\tfs=\tys$. In this case, we find the object
  $n_{\tx_i}$ corresponding to each argument $\tx_i$ passed into the
  constructor (it is the third component of the relationship whose
  label is $\tx_i$), and construct a new relationship with an empty
  property set $(f, n_{\tC}, \Nods(\tx_i).3, \emptyset)$.
\item If $C$ extends some other class $D$, then the first $k$
  arguments (denoted as $\txs_1^k$) to the constructor will be used to
  initialize $D$, while the rest will be used to initialize $C$'s
  fields. To construct the full set of new relationships, we
  recursively call $\fs$ for $D$ with the first $k$ arguments, and
  then augment the resulting set with the initializations for $C$'s
  fields, calculated as in the base case.
\end{itemize}

\subsection{Implementation}
\label{sec:implementation}

We now describe the implementation of \Tool by considering the
execution of the \Cypher query in
Figure~\ref{figure:create_ds_example_match_specific_reachability}.
The execution can be divided into the following 4 steps.

\MyPara{(1)~Pre-processing queries}
Semantically, the query aims to find patterns that contain instances
referring to other instances that are 2 hops away, referenced through
fields named \Code{left} or \Code{right} and those which contain a
primitive \Code{int} field named \Code{value} with value 1. The first
step involves processing the query format string to expand the positional
arguments using the expansion described in Table~\ref{fig:positional:arguments}.
The positional arguments \Code{\$1} and \Code{@1} are expanded into a unique
identifier and the fully qualified class name for \Code{bTree.root},
respectively.
Any function that maps objects bijectively to a set comprised of
either, one of the \Java primitive data types or
\Code{java.lang.String} elements can be used to generate the unique
identifier in the expansion of \Code{\$}. This constraint is based
on the design of our property graph model where \Java primitive and
\Code{String} fields are embedded as node properties. We use the
hashcode computed using the \Code{identityHashCode(Object
  obj)}~\cite{Java_Hashcode} method from the
\Code{java.lang.}\-\Code{System} package as the unique identifier
in our implementation. This function is bijective except for some
pathlogical scenarios~\cite{Java_Hashcode_Caveat}, and can be easily
replaced with another function in the future.

\MyPara{(2)~Triggering subgraph identification}
The next step is the identification of a subgraph of objects on the
\JVM heap memory, of relevance to the query. Since a significant
number of objects in the heap may be irrelevant to the query, it is
more efficient to index once and store a subgraph of the complete
object graph than to re-traverse it for every additional clause in the
query. We store the subgraph in semantically corresponding lightweight
C++ classes (for e.g., a \Java \Code{class} definition is stored in
C++ \Code{ClassInfo} class that stores information about the class
name, its field names, their types and modifiers).

We use a native agent developed using the \JVMTI\cite{jvmtiOracle}
framework to identify the subgraph. Native agents can be triggered
using \JVMTI events based on certain actions from within a \Java
application.  The general approach is to perform an action in the
\Java domain of \Tool after pre-processing the query that triggers an
event inside \JVMTI.  The callback provided by \JVMTI to service the
event can then be used as the entry point to the native agent. We use
the exception event, generated when an exception (any instance of
\Code{java.lang.Throwable}) is thrown by a \Java method to trigger
the native agent. We created a dummy exception class
\Code{GraphTriggerException}.  This exception is thrown from a dummy
method (\Code{setupGraph}) after the query pre-processing step. In
the \JVMTI callback for handling exception events, we monitor if the
signature of the method from which this exception is thrown matches
that of \Code{setupGraph}. If a match is found then the native agent
proceeds with identification of the subgraph.

\MyPara{(3)~Identifying subgraphs}
\Tool uses \JVMTI to identify relations between objects and their
field properties (name, value, type, etc.).
The native agent is implemented in the C++ programming language. The
following are the major steps involved in identifying the subgraph.

\MyParaTwo{Tag zero} \JVMTI callbacks use \emph{tags} which are
primitive long types to refer to objects on the heap. These tags can
be modified inside certain \JVMTI callbacks and is the preferred way
of identifying and tracking relations across objects. However, these
tags can also be modified by \JVMTI internal processes. Therefore, as
an initialization step, we iterate through all objects on the heap and
set their tags to 0. Optionally, our optimization--\emph{Force
Garbage Collection}--can be enabled to force a GC event prior to
tagging to reduce the total number of objects on the heap
consequently, reducing the overhead incurred by all the following
\JVMTI callbacks.

\MyParaTwo{Loaded classes}
The next step involves identifying the
types of the objects potentially relevant to the query. This helps
limit the number of objects to be considered for inclusion into the
subgraph. \Tool provides an optimization \emph{whitelist} which can be
used to flag certain user specified classes whose instances are
guaranteed to be included in the subgraph. In addition, \Tool also
provides a \emph{blacklist} for specifying classes whose instances are
to be definitively excluded from the subgraph. Furthermore, in
addition to each object being referred to by tags, \JVMTI framework
also uses tags to identify classes. By initializing all objects to 0
in the \emph{Tag zero} step, we also initialize the tags of all
classes to 0 (instances of \Code{java.lang.Class}). When a class is
blacklisted, it remains untagged (tag=0) and hence its instances will
not be reported in any of the following \JVMTI callbacks since all
\JVMTI callbacks support filters to filter out untagged objects and
classes.

\MyParaTwo{Iterate heap assign unique tag} In this step we assign a
unique tag to every instance of every tagged class. We also allocate
memory to certain bookeeping C++ classes required for storing
information about the subgraph. These unique tags help identify
relations between objects in the next step.

\MyParaTwo{Follow references}
This \JVMTI callback traverses the
object graph on the \JVM heap. It first reports the referrer and
referee instances followed by the primitive fields, String fields and
array primitive fields of the referrer instance before doing the like
for the referee instance. We once again limit the objects reported by
this callback by applying filters ensuring that the reported objects
as well as their classes are tagged. By default, if no root object is
specified, the traversal is started from a set of system classes, \JNI
globals and other objects used as roots for garbage collection. The
optimization \emph{Fix Root Objects} can be used to start the
traversal from the specified object. This reduces the overhead of
traversing paths irrelevant to the specified query.

\MyParaTwo{Write graph to CSV} This step applies to \ToolNeo
exclusively. In this step, we serialize the subgraph into CSV files
such that they can be batch imported into a \Neo database.

\MyPara{(4)~Executing the query} For the \ToolNeo implementation, the
exported CSV files are loaded into a \Neo database and the given query
is executed using \Neo's \Cypher engine. For queries involving the
return of a primitive/String fields (stored as properties of \Neo
nodes), the result of \Neo's \Cypher engine is the result of the
specified query. For queries that involve the return of an object, we
use the semantically equivalent \Neo node returned by \Neo's \Cypher
engine to obtain the unique identifier.
Followed by this, we use the \JNI framework to retrieve the
corresponding instance from the \JVM heap.

For the \ToolMemory implementation, we used the
\Antlr~\cite{Parr_Antlr4,Antlr4} parser generator to generate the
parser and visitor for the \openCypher query language~\cite{Alastair_openCypher}. We use the visitor
pattern to deduce the semantics and execute the query. In this
implementation, there is no overhead of writing the subgraph to CSV or
setting up and creating a \Neo graph mirroring the state of the \JVM's
heap. The query is executed directly on the subgraph and like former,
\JNI is used to report the result back to \Java.

\subsection{Optimizations}
\label{sec:optimizations}
%
%
To improve the performance of \Tool, we introduced 3 optimizations,
Whitelist (\xWhitelist), Force Garbage Collection (FGC), and Fix Root
Objects (\xTraversePassedObjects).

\MyPara{Whitelist (\xWhitelist)}
Limits the size of the subgraph by specifying the type of instances to
be definitively included (instances reachable from the specified
instance types under transitive closure are also included).

\MyPara{Force garbage collection (\textbf{FGC})}
Force a garbage collection event to reduce the number of objects on
the \JVM's heap before performing the steps to identify the
subgraph. This decreases the overhead incurred during the \JVMTI
callbacks.

\MyPara{Fix root objects (\textbf{FRO})}
Limits \JVMTI \Code{\FollowReferences} 
callbacks to reporting instances that are transitively reachable from 
the provided root object. The root object is passed as an argument 
to the query \API call (i.e., \boundedq).

\section{Evaluation}
\label{sec:eval}

We evaluated \Tool in two parts: (1)~by rewriting existing assertion
statements available in tests in
open-source projects, and (2)~by implementing methods from \Java data
structure libraries. The first part demonstrates the robustness of our
system and ease of its integration with large open-source projects
while the second part describes its expressive power over a purely
imperative approach.  Most of the selected projects are supported by
large software organizations, such as Apache or Google.
This section describes the chosen subjects and our findings.
The \Tool queries were benchmarked on a 64-bit Ubuntu 18.04.1 desktop
with a 11th Gen Intel(R) Core(TM) i7-8700 @ 3.20GHz and 64GB
RAM. We use Java \JavaVersion and \Neo \NeoVersion for
all experiments.

\begin{table*}%
  \TableFontSize%
  \caption{\PerformanceCaption}%
  \vspace{-3mm}%
  \label{tab:performance}%
  \begin{tabular*}{\textwidth}{@{\extracolsep{\fill}}llrrrrrrrrrrc@{}}%
    \toprule%
    \multirow{3}{*}{\Project} &%
    \multirow{3}{*}{\SHA} &%
    \multirow{3}{*}{\textbf{LOC}} &%
    \multirow{3}{*}{\textbf{\#Assert}} &%
    \multicolumn{5}{c}{\textbf{\ToolNeo}} &%
    \multicolumn{3}{c}{\textbf{\ToolMemory}} &%
    \\%
    \cline{5-9}\cline{10-12}%
    &%
    &%
    &%
    &%
    &%
    \multicolumn{4}{c}{+\Whitelist} &%
    &%
    \multicolumn{2}{c}{+\Whitelist} &%
    \ReductionAntlrFGCNeoAll %
    \\%
    \cline{6-9} \cline{11-12}%
    \multicolumn{4}{c}{} & %
    \Naive & %
    \multicolumn{1}{c}{-} &%
    \TraversePassedObjects &%
    \ForceGC &%
    \All &%
    \AntlrNaive &%
    \multicolumn{1}{c}{-} &%
    \AntlrForceGC &%
    \textbf{[\%]} %
    \\%
    \midrule%
    \assertjDB &%
    \href{\assertjDBURL}{\urlsymbol} \assertjDBSHA &%
    \assertjDBLOC &%
    \assertjDBNumAssertions &%
    \UseMacro{assertjDBNoneAvg} &%
    \UseMacro{assertjDBWhitelistAvg} &%
    \UseMacro{assertjDBTraversePassedObjectsAvg} &%
    \UseMacro{assertjDBForceGCAvg} &%
    \UseMacro{assertjDBAllAvg} &%
    \UseMacro{assertjDBAntlrNoneAvg} &%
    \UseMacro{assertjDBAntlrWhitelistAvg} &%
    \UseMacro{assertjDBAntlrForceGCAvg} &%
    \UseMacro{assertjDBAntlrFGCNeoAll}%
    \\%
    \commonsCli &%
    \href{\commonsCliURL}{\urlsymbol} \commonsCliSHA &%
    \commonsCliLOC &%
    \commonsCliNumAssertions &%
    \UseMacro{commonsCliNoneAvg} &%
    \UseMacro{commonsCliWhitelistAvg} &%
    \UseMacro{commonsCliTraversePassedObjectsAvg} &%
    \UseMacro{commonsCliForceGCAvg} &%
    \UseMacro{commonsCliAllAvg} &%
    \UseMacro{commonsCliAntlrNoneAvg} &%
    \UseMacro{commonsCliAntlrWhitelistAvg} &%
    \UseMacro{commonsCliAntlrForceGCAvg} &%
    \UseMacro{commonsCliAntlrFGCNeoAll}%
    \\%
    \commonsCsv &%
    \href{\commonsCsvURL}{\urlsymbol} \commonsCsvSHA &%
    \commonsCsvLOC &%
    \commonsCsvNumAssertions &%
    \UseMacro{commonsCsvNoneAvg} &%
    \UseMacro{commonsCsvWhitelistAvg} &%
    \UseMacro{commonsCsvTraversePassedObjectsAvg} &%
    \UseMacro{commonsCsvForceGCAvg} &%
    \UseMacro{commonsCsvAllAvg} &%
    \UseMacro{commonsCsvAntlrNoneAvg} &%
    \UseMacro{commonsCsvAntlrWhitelistAvg} &%
    \UseMacro{commonsCsvAntlrForceGCAvg} &%
    \UseMacro{commonsCsvAntlrFGCNeoAll}%
    \\%
    \commonsLang &%
    \href{\commonsLangURL}{\urlsymbol} \commonsLangSHA &%
    \commonsLangLOC &%
    \commonsLangNumAssertions &%
    \UseMacro{commonsLangNoneAvg} &%
    \UseMacro{commonsLangWhitelistAvg} &%
    \UseMacro{commonsLangTraversePassedObjectsAvg} &%
    \UseMacro{commonsLangForceGCAvg} &%
    \UseMacro{commonsLangAllAvg} &%
    \UseMacro{commonsLangAntlrNoneAvg} &%
    \UseMacro{commonsLangAntlrWhitelistAvg} &%
    \UseMacro{commonsLangAntlrForceGCAvg} &%
    \UseMacro{commonsLangAntlrFGCNeoAll}%
    \\%
    \commonsGeometry &%
    \href{\commonsGeometryURL}{\urlsymbol} \commonsGeometrySHA &%
    \commonsGeometryLOC &%
    \commonsGeometryNumAssertions &%
    \UseMacro{commonsGeometryNoneAvg} &%
    \UseMacro{commonsGeometryWhitelistAvg} &%
    \UseMacro{commonsGeometryTraversePassedObjectsAvg} &%
    \UseMacro{commonsGeometryForceGCAvg} &%
    \UseMacro{commonsGeometryAllAvg} &%
    \UseMacro{commonsGeometryAntlrNoneAvg} &%
    \UseMacro{commonsGeometryAntlrWhitelistAvg} &%
    \UseMacro{commonsGeometryAntlrForceGCAvg} &%
    \UseMacro{commonsGeometryAntlrFGCNeoAll}%
    \\%
    \guava &%
    \href{\guavaURL}{\urlsymbol} \guavaSHA &%
    \guavaLOC &%
    \guavaNumAssertions &%
    \UseMacro{guavaNoneAvg} &%
    \UseMacro{guavaWhitelistAvg} &%
    \UseMacro{guavaTraversePassedObjectsAvg} &%
    \UseMacro{guavaForceGCAvg} &%
    \UseMacro{guavaAllAvg} &%
    \UseMacro{guavaAntlrNoneAvg} &%
    \UseMacro{guavaAntlrWhitelistAvg} &%
    \UseMacro{guavaAntlrForceGCAvg} &%
    \UseMacro{guavaAntlrFGCNeoAll}%
    \\%
    \jfreeChart &%
    \href{\jfreeChartURL}{\urlsymbol} \jfreeChartSHA &%
    \jfreeChartLOC &%
    \jfreeChartNumAssertions &%
    \UseMacro{jfreeChartNoneAvg} &%
    \UseMacro{jfreeChartWhitelistAvg} &%
    \UseMacro{jfreeChartTraversePassedObjectsAvg} &%
    \UseMacro{jfreeChartForceGCAvg} &%
    \UseMacro{jfreeChartAllAvg} &%
    \UseMacro{jfreeChartAntlrNoneAvg} &%
    \UseMacro{jfreeChartAntlrWhitelistAvg} &%
    \UseMacro{jfreeChartAntlrForceGCAvg} &%
    \UseMacro{jfreeChartAntlrFGCNeoAll}%
    \\%
    \jsonJava &%
    \href{\jsonJavaURL}{\urlsymbol} \jsonJavaSHA &%
    \jsonJavaLOC &%
    \jsonJavaNumAssertions &%
    \UseMacro{jsonJavaNoneAvg} &%
    \UseMacro{jsonJavaWhitelistAvg} &%
    \UseMacro{jsonJavaTraversePassedObjectsAvg} &%
    \UseMacro{jsonJavaForceGCAvg} &%
    \UseMacro{jsonJavaAllAvg} &%
    \UseMacro{jsonJavaAntlrNoneAvg} &%
    \UseMacro{jsonJavaAntlrWhitelistAvg} &%
    \UseMacro{jsonJavaAntlrForceGCAvg} &%
    \UseMacro{jsonJavaAntlrFGCNeoAll}%
    \\%
    \tcases &%
    \href{\tcasesURL}{\urlsymbol} \tcasesSHA &%
    \tcasesLOC &%
    \tcasesNumAssertions &%
    \UseMacro{tcasesNoneAvg} &%
    \UseMacro{tcasesWhitelistAvg} &%
    \UseMacro{tcasesTraversePassedObjectsAvg} &%
    \UseMacro{tcasesForceGCAvg} &%
    \UseMacro{tcasesAllAvg} &%
    \UseMacro{tcasesAntlrNoneAvg} &%
    \UseMacro{tcasesAntlrWhitelistAvg} &%
    \UseMacro{tcasesAntlrForceGCAvg} &%
    \UseMacro{tcasesAntlrFGCNeoAll}%
    \\%
    \zipFourj &%
    \href{\zipFourjURL}{\urlsymbol} \zipFourjSHA &%
    \zipFourjLOC &%
    \zipFourjNumAssertions &%
    \UseMacro{zipFourjNoneAvg} &%
    \UseMacro{zipFourjWhitelistAvg} &%
    \UseMacro{zipFourjTraversePassedObjectsAvg} &%
    \UseMacro{zipFourjForceGCAvg} &%
    \UseMacro{zipFourjAllAvg} &%
    \UseMacro{zipFourjAntlrNoneAvg} &%
    \UseMacro{zipFourjAntlrWhitelistAvg} &%
    \UseMacro{zipFourjAntlrForceGCAvg} &%
    \UseMacro{zipFourjAntlrFGCNeoAll}%
    \\%
    \midrule%
    \xAvg &\multicolumn{1}{c}{\multirow{2}{*}{-}}%
    &\multicolumn{1}{c}{\multirow{2}{*}{-}}%
    &\multicolumn{1}{r}{\multirow{2}{*}{-}}%
    &%
    \NaiveNeoAverage &%
    \WlNeoAverage &%
    \FRONeoAverage &%
    \FGCNeoAverage &%
    \AllNeoAverage &%
    \NaiveAntlrAverage &%
    \WlAntlrAverage &%
    \FGCAntlrAverage & \multicolumn{1}{c}{\multirow{2}{*}{-}}%
    \\%
    \xTotal &%
    &%
    &%
    &%
    \NaiveNeoTotal &%
    \WlNeoTotal &%
    \FRONeoTotal &%
    \FGCNeoTotal &%
    \AllNeoTotal &%
    \NaiveAntlrTotal &%
    \WlAntlrTotal &%
    \FGCAntlrTotal &%
    \\%
    \bottomrule%
  \end{tabular*}%
\end{table*}%

\subsection{Results}

This section describes our findings.

\MyPara{Re-written assertions}
The selected projects, their lines of code (LOC) and \Tool query
execution times under different introduced optimizations are given in
Table~\ref{tab:performance}. All executions are averaged over 50 runs
except for the \Naive case that is averaged over 3 runs due to long
execution time.
The reported times are all end-to-end and limited to the scope of
\Tool API methods (includes the object graph construction time, the
query time itself and in case of \ToolNeo, serialization and clearing
of \Neo database). We do not consider the total test time, as long
running tests would then mask the actual cost of queries.
All times are in milliseconds.

Columns 5 through 9 report the query times for \ToolNeo prototype
whereas those for \ToolMemory are given in columns 10 through 12. The
final column shows the speedup of \ToolMemory over \ToolNeo.
For \ToolNeo and \ToolMemory we show time with different optimization
levels, which were described in Section~\ref{sec:optimizations}.  To
compute the speedup, we use \ToolNeo:\xAll and
\ToolMemory:\xForceGC.
The last two rows in the table show the average (\xAvg) and the total 
time (\xTotal) across all the assertions from all projects.

\ToolNeo:\xNaive is not usable since this attempts to include every
object on the heap into the graph; there is even a case when the
entire run crashed as the VM ran out of memory (\commonsGeometry).
Each individual optimization provides substantial reduction in the
execution time over \xNaive. \xForceGC provides the biggest reduction
as it dramatically reduces the number of objects to be translated.
Finally, combining all the optimizations (\ToolNeo:\xAll) together
gives the best performance in most cases.
Speedup of \ToolNeo:\xAll over \ToolNeo:\xNaive is up to 94\%.
Looking at the third column, we find that \ToolMemory:\xNaive is
substantially faster than even the most optimal \ToolNeo.
Furthermore, using \xForceGC improves the times of \ToolMemory by half.

In summary, we find that our optimizations are effective.
Furthermore, \ToolMemory is, at the moment, better suited for writing
program statements due to its low cost.  However, we still see
\ToolNeo being very much usable in a debugging environment, where
moderate overhead with a large number of features provided by \Neo can
be effectively used.%
\begin{figure}[bp]
  \centering
  \def\svgwidth{\columnwidth}
  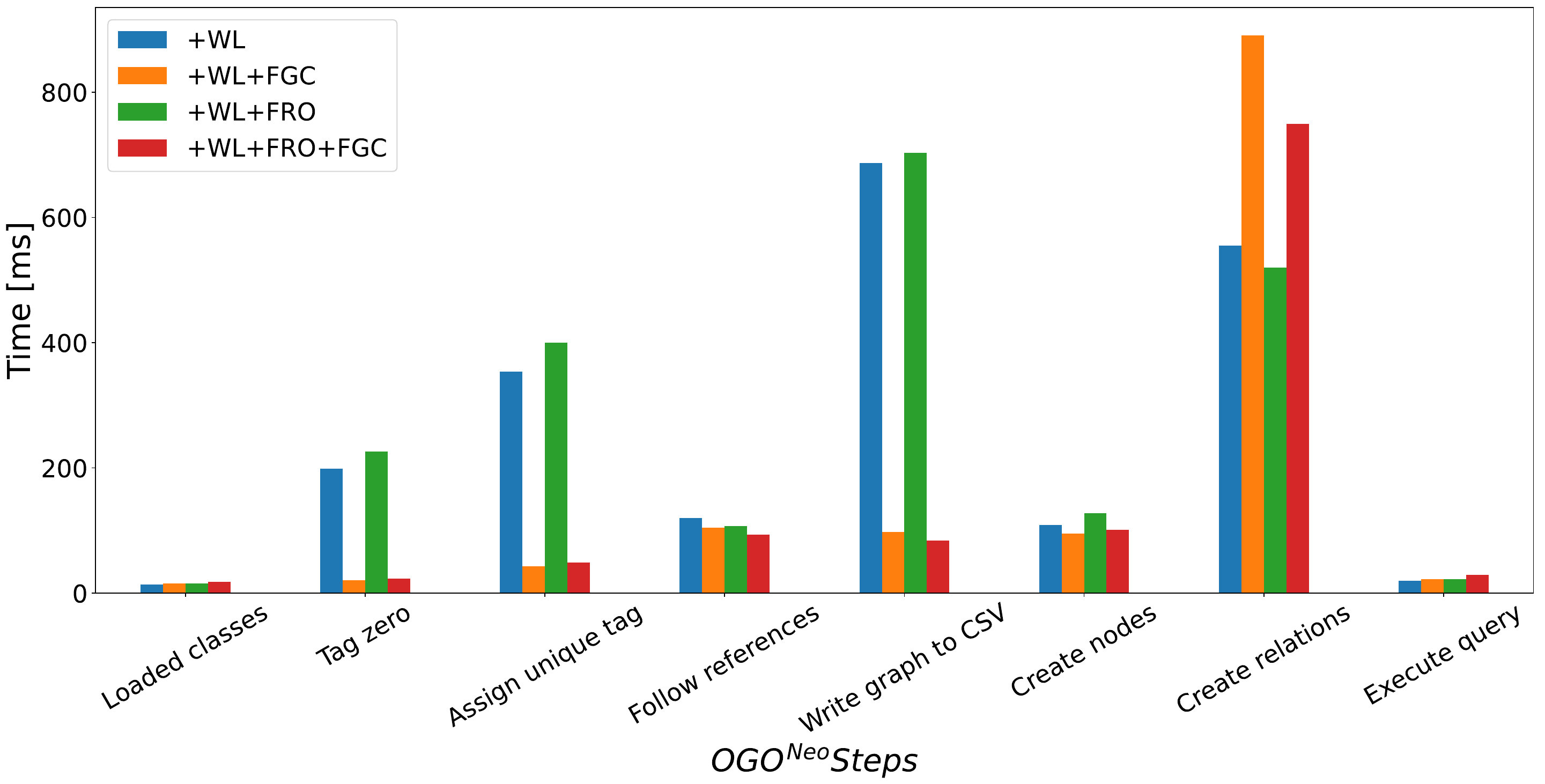
  \caption{\FigOGOStepBreakdownCaption}
  \label{figure:step_breakdown}
\end{figure}

\MyPara{Implementing library methods}
We re-wrote methods of classes from \guava, the Java Collections
Framework (\jcf) and the Java Universal Network/Graph Framework (\jung)
using \Tool. Table~\ref{tab:loc_noc} shows the average and total lines
of code (LOC) and number of characters (NOC) for implementing the
selected methods using \Tool and a purely imperative approach (which is already
available in those libraries). We see
that on average, \Tool requires 5 and 4 times as less LOC and NOC than
its counterpart.

To confirm the validity of our implementation, we executed
  in total \guavaUnitTests test methods for \guava, \jungUnitTests for
  \jung and \jcfUnitTests for \jcf. We executed all the test methods 
  for \guava and \jung and only the test methods (present in \Code{java.util}
  package) for modified classes for \jcf. While \guava and \jung both used
  JUnit as the test runner, \jcf used JTReg.  We setup the tests by
  replacing the imperative implementation of the selected methods with
  the \Tool implementation and then executing the tests.

\begin{table}[t]%
  \TableFontSize%
  \caption{\UseMacro{TableLocNocCaption}}%
  \vspace{-4mm}%
  \label{tab:loc_noc}%
\begin{tabular*}{0.45\textwidth}{@{\extracolsep{\fill}}@{}l@{\hskip 0.2em}l @{\hskip 0em} rr rr@{}}%
\toprule%
\multicolumn{2}{c}{\multirow{2}{*}{\UseMacro{TableLocNocClass}}} & \multicolumn{2}{c}{\UseMacro{TableLocNocOgo}} & \multicolumn{2}{c}{\UseMacro{TableLocNocImp}} \\%
\multicolumn{2}{c}{} &  \UseMacro{TableLocNocLoc} &  \UseMacro{TableLocNocNoc} &  \UseMacro{TableLocNocLoc} &  \UseMacro{TableLocNocNoc} \\%
\midrule%
\multirow{3}{*}{\rotatebox[origin=c]{270}{\UseMacro{TableLocNocGuava}}}%
& \UseMacro{TableLocNocArrayTable}%
& \UseMacro{TableLocNocArrayTableOgoLoc}%
& \UseMacro{TableLocNocArrayTableOgoNoc}%
& \UseMacro{TableLocNocArrayTableImpLoc}%
& \UseMacro{TableLocNocArrayTableImpNoc}%
\\%
& \UseMacro{TableLocNocHashBiMap}%
& \UseMacro{TableLocNocHashBiMapOgoLoc}%
& \UseMacro{TableLocNocHashBiMapOgoNoc}%
& \UseMacro{TableLocNocHashBiMapImpLoc}%
& \UseMacro{TableLocNocHashBiMapImpNoc}%
\\%
& \UseMacro{TableLocNocLinkedListMultiMap}%
& \UseMacro{TableLocNocLinkedListMultiMapOgoLoc}%
& \UseMacro{TableLocNocLinkedListMultiMapOgoNoc}%
& \UseMacro{TableLocNocLinkedListMultiMapImpLoc}%
& \UseMacro{TableLocNocLinkedListMultiMapImpNoc}%
\\%
\midrule%
\multirow{3}{*}{\rotatebox[origin=c]{270}{\UseMacro{TableLocNocJUNG}}}%
& \UseMacro{TableLocNocSparseGraph}%
& \UseMacro{TableLocNocSparseGraphOgoLoc}%
& \UseMacro{TableLocNocSparseGraphOgoNoc}%
& \UseMacro{TableLocNocSparseGraphImpLoc}%
& \UseMacro{TableLocNocSparseGraphImpNoc}%
\\%
& \UseMacro{TableLocNocUndirectedSparseGraph}%
& \UseMacro{TableLocNocUndirectedSparseGraphOgoLoc}%
& \UseMacro{TableLocNocUndirectedSparseGraphOgoNoc}%
& \UseMacro{TableLocNocUndirectedSparseGraphImpLoc}%
& \UseMacro{TableLocNocUndirectedSparseGraphImpNoc}%
\\%
& \UseMacro{TableLocNocDirectedSparseGraph}%
& \UseMacro{TableLocNocDirectedSparseGraphOgoLoc}%
& \UseMacro{TableLocNocDirectedSparseGraphOgoNoc}%
& \UseMacro{TableLocNocDirectedSparseGraphImpLoc}%
& \UseMacro{TableLocNocDirectedSparseGraphImpNoc}%
\\%
\midrule%
\multirow{5}{*}{\rotatebox[origin=c]{270}{\UseMacro{TableLocNocJCF}}}%
& \UseMacro{TableLocNocArrayList}%
& \UseMacro{TableLocNocArrayListOgoLoc}%
& \UseMacro{TableLocNocArrayListOgoNoc}%
& \UseMacro{TableLocNocArrayListImpLoc}%
& \UseMacro{TableLocNocArrayListImpNoc}%
\\%
& \UseMacro{TableLocNocHashMap}%
& \UseMacro{TableLocNocHashMapOgoLoc}%
& \UseMacro{TableLocNocHashMapOgoNoc}%
& \UseMacro{TableLocNocHashMapImpLoc}%
& \UseMacro{TableLocNocHashMapImpNoc}%
\\%
& \UseMacro{TableLocNocLinkedList}%
& \UseMacro{TableLocNocLinkedListOgoLoc}%
& \UseMacro{TableLocNocLinkedListOgoNoc}%
& \UseMacro{TableLocNocLinkedListImpLoc}%
& \UseMacro{TableLocNocLinkedListImpNoc}%
\\%
& \UseMacro{TableLocNocArrayDeque}%
& \UseMacro{TableLocNocArrayDequeOgoLoc}%
& \UseMacro{TableLocNocArrayDequeOgoNoc}%
& \UseMacro{TableLocNocArrayDequeImpLoc}%
& \UseMacro{TableLocNocArrayDequeImpNoc}%
\\%
& \UseMacro{TableLocNocVector}%
& \UseMacro{TableLocNocVectorOgoLoc}%
& \UseMacro{TableLocNocVectorOgoNoc}%
& \UseMacro{TableLocNocVectorImpLoc}%
& \UseMacro{TableLocNocVectorImpNoc}%
\\%
\midrule%
\UseMacro{TableLocNocAvg}%
& \multicolumn{1}{c}{-}%
& \UseMacro{TableLocNocAvgValueOgoLoc}%
& \UseMacro{TableLocNocAvgValueOgoNoc}%
& \UseMacro{TableLocNocAvgValueImpLoc}%
& \UseMacro{TableLocNocAvgValueImpNoc}%
\\%
\UseMacro{TableLocNocSum}%
& \multicolumn{1}{c}{-}%
& \UseMacro{TableLocNocSumValueOgoLoc}%
& \UseMacro{TableLocNocSumValueOgoNoc}%
& \UseMacro{TableLocNocSumValueImpLoc}%
& \UseMacro{TableLocNocSumValueImpNoc}%
\\%
\bottomrule%
\end{tabular*}%
\end{table}%

\begin{table*}[t]%
\TableFontSize%
\caption{\UseMacro{TableDataStructureEvalCaption}}%
\label{tab:table_data_structure}%
\begin{tabular*}{\textwidth}{@{\extracolsep{\fill}}lrrrrrrrr}%
\toprule%
\multicolumn{1}{c}{\multirow{3}{*}{\UseMacro{TableDataStructureEvalClass}}} & \multicolumn{4}{c}{\UseMacro{TableDataStructureEvalOGONeo}} & \multicolumn{4}{c}{\UseMacro{TableDataStructureEvalOGOMem}}%
\\%
\multicolumn{1}{c}{} & \multicolumn{4}{c}{ \UseMacro{TableDataStructureEvalWL}} & \multicolumn{4}{c}{ \UseMacro{TableDataStructureEvalWL}}%
\\%
\cline{2-5} \cline{6-9}%
\multicolumn{1}{c}{} & \multicolumn{1}{c}{-} &  \UseMacro{TableDataStructureEvalFRO} &  \UseMacro{TableDataStructureEvalFGC} &  \UseMacro{TableDataStructureEvalFROFGC} & \multicolumn{1}{c}{-} &  \UseMacro{TableDataStructureEvalFRO} &  \UseMacro{TableDataStructureEvalFGC} &  \UseMacro{TableDataStructureEvalFROFGC}%
\\%
\midrule%
\multirow{1}{*}{\UseMacro{TableDataStructureEvalVector}}
& \UseMacro{TableDataStructureEvalNeoVectorTwoHundredWLAvg}%
& \UseMacro{TableDataStructureEvalNeoVectorTwoHundredWLFGCAvg}%
& \UseMacro{TableDataStructureEvalNeoVectorTwoHundredWLFROAvg}%
& \UseMacro{TableDataStructureEvalNeoVectorTwoHundredWLFROFGCAvg}%
& \UseMacro{TableDataStructureEvalMemVectorTwoHundredWLAvg}%
& \UseMacro{TableDataStructureEvalMemVectorTwoHundredWLFGCAvg}%
& \UseMacro{TableDataStructureEvalMemVectorTwoHundredWLFROAvg}%
& \UseMacro{TableDataStructureEvalMemVectorTwoHundredWLFROFGCAvg}%
\\%
\multirow{1}{*}{\UseMacro{TableDataStructureEvalUndirectedSparseGraph}}
& \UseMacro{TableDataStructureEvalNeoUndirectedSparseGraphTwoHundredWLAvg}%
& \UseMacro{TableDataStructureEvalNeoUndirectedSparseGraphTwoHundredWLFGCAvg}%
& \UseMacro{TableDataStructureEvalNeoUndirectedSparseGraphTwoHundredWLFROAvg}%
& \UseMacro{TableDataStructureEvalNeoUndirectedSparseGraphTwoHundredWLFROFGCAvg}%
& \UseMacro{TableDataStructureEvalMemUndirectedSparseGraphTwoHundredWLAvg}%
& \UseMacro{TableDataStructureEvalMemUndirectedSparseGraphTwoHundredWLFGCAvg}%
& \UseMacro{TableDataStructureEvalMemUndirectedSparseGraphTwoHundredWLFROAvg}%
& \UseMacro{TableDataStructureEvalMemUndirectedSparseGraphTwoHundredWLFROFGCAvg}%
\\%
\multirow{1}{*}{\UseMacro{TableDataStructureEvalHashBiMap}}
& \UseMacro{TableDataStructureEvalNeoHashBiMapTwoHundredWLAvg}%
& \UseMacro{TableDataStructureEvalNeoHashBiMapTwoHundredWLFGCAvg}%
& \UseMacro{TableDataStructureEvalNeoHashBiMapTwoHundredWLFROAvg}%
& \UseMacro{TableDataStructureEvalNeoHashBiMapTwoHundredWLFROFGCAvg}%
& \UseMacro{TableDataStructureEvalMemHashBiMapTwoHundredWLAvg}%
& \UseMacro{TableDataStructureEvalMemHashBiMapTwoHundredWLFGCAvg}%
& \UseMacro{TableDataStructureEvalMemHashBiMapTwoHundredWLFROAvg}%
& \UseMacro{TableDataStructureEvalMemHashBiMapTwoHundredWLFROFGCAvg}%
\\%
\multirow{1}{*}{\UseMacro{TableDataStructureEvalHashMap}}
& \UseMacro{TableDataStructureEvalNeoHashMapTwoHundredWLAvg}%
& \UseMacro{TableDataStructureEvalNeoHashMapTwoHundredWLFGCAvg}%
& \UseMacro{TableDataStructureEvalNeoHashMapTwoHundredWLFROAvg}%
& \UseMacro{TableDataStructureEvalNeoHashMapTwoHundredWLFROFGCAvg}%
& \UseMacro{TableDataStructureEvalMemHashMapTwoHundredWLAvg}%
& \UseMacro{TableDataStructureEvalMemHashMapTwoHundredWLFGCAvg}%
& \UseMacro{TableDataStructureEvalMemHashMapTwoHundredWLFROAvg}%
& \UseMacro{TableDataStructureEvalMemHashMapTwoHundredWLFROFGCAvg}%
\\%
\multirow{1}{*}{\UseMacro{TableDataStructureEvalSparseGraph}}
& \UseMacro{TableDataStructureEvalNeoSparseGraphTwoHundredWLAvg}%
& \UseMacro{TableDataStructureEvalNeoSparseGraphTwoHundredWLFGCAvg}%
& \UseMacro{TableDataStructureEvalNeoSparseGraphTwoHundredWLFROAvg}%
& \UseMacro{TableDataStructureEvalNeoSparseGraphTwoHundredWLFROFGCAvg}%
& \UseMacro{TableDataStructureEvalMemSparseGraphTwoHundredWLAvg}%
& \UseMacro{TableDataStructureEvalMemSparseGraphTwoHundredWLFGCAvg}%
& \UseMacro{TableDataStructureEvalMemSparseGraphTwoHundredWLFROAvg}%
& \UseMacro{TableDataStructureEvalMemSparseGraphTwoHundredWLFROFGCAvg}%
\\%
\multirow{1}{*}{\UseMacro{TableDataStructureEvalArrayDeque}}
& \UseMacro{TableDataStructureEvalNeoArrayDequeTwoHundredWLAvg}%
& \UseMacro{TableDataStructureEvalNeoArrayDequeTwoHundredWLFGCAvg}%
& \UseMacro{TableDataStructureEvalNeoArrayDequeTwoHundredWLFROAvg}%
& \UseMacro{TableDataStructureEvalNeoArrayDequeTwoHundredWLFROFGCAvg}%
& \UseMacro{TableDataStructureEvalMemArrayDequeTwoHundredWLAvg}%
& \UseMacro{TableDataStructureEvalMemArrayDequeTwoHundredWLFGCAvg}%
& \UseMacro{TableDataStructureEvalMemArrayDequeTwoHundredWLFROAvg}%
& \UseMacro{TableDataStructureEvalMemArrayDequeTwoHundredWLFROFGCAvg}%
\\%
\multirow{1}{*}{\UseMacro{TableDataStructureEvalArrayList}}
& \UseMacro{TableDataStructureEvalNeoArrayListTwoHundredWLAvg}%
& \UseMacro{TableDataStructureEvalNeoArrayListTwoHundredWLFGCAvg}%
& \UseMacro{TableDataStructureEvalNeoArrayListTwoHundredWLFROAvg}%
& \UseMacro{TableDataStructureEvalNeoArrayListTwoHundredWLFROFGCAvg}%
& \UseMacro{TableDataStructureEvalMemArrayListTwoHundredWLAvg}%
& \UseMacro{TableDataStructureEvalMemArrayListTwoHundredWLFGCAvg}%
& \UseMacro{TableDataStructureEvalMemArrayListTwoHundredWLFROAvg}%
& \UseMacro{TableDataStructureEvalMemArrayListTwoHundredWLFROFGCAvg}%
\\%
\multirow{1}{*}{\UseMacro{TableDataStructureEvalArrayTable}}
& \UseMacro{TableDataStructureEvalNeoArrayTableTwoHundredWLAvg}%
& \UseMacro{TableDataStructureEvalNeoArrayTableTwoHundredWLFGCAvg}%
& \UseMacro{TableDataStructureEvalNeoArrayTableTwoHundredWLFROAvg}%
& \UseMacro{TableDataStructureEvalNeoArrayTableTwoHundredWLFROFGCAvg}%
& \UseMacro{TableDataStructureEvalMemArrayTableTwoHundredWLAvg}%
& \UseMacro{TableDataStructureEvalMemArrayTableTwoHundredWLFGCAvg}%
& \UseMacro{TableDataStructureEvalMemArrayTableTwoHundredWLFROAvg}%
& \UseMacro{TableDataStructureEvalMemArrayTableTwoHundredWLFROFGCAvg}%
\\%
\multirow{1}{*}{\UseMacro{TableDataStructureEvalLinkedList}}
& \UseMacro{TableDataStructureEvalNeoLinkedListTwoHundredWLAvg}%
& \UseMacro{TableDataStructureEvalNeoLinkedListTwoHundredWLFGCAvg}%
& \UseMacro{TableDataStructureEvalNeoLinkedListTwoHundredWLFROAvg}%
& \UseMacro{TableDataStructureEvalNeoLinkedListTwoHundredWLFROFGCAvg}%
& \UseMacro{TableDataStructureEvalMemLinkedListTwoHundredWLAvg}%
& \UseMacro{TableDataStructureEvalMemLinkedListTwoHundredWLFGCAvg}%
& \UseMacro{TableDataStructureEvalMemLinkedListTwoHundredWLFROAvg}%
& \UseMacro{TableDataStructureEvalMemLinkedListTwoHundredWLFROFGCAvg}%
\\%
\multirow{1}{*}{\UseMacro{TableDataStructureEvalDirectedSparseGraph}}
& \UseMacro{TableDataStructureEvalNeoDirectedSparseGraphTwoHundredWLAvg}%
& \UseMacro{TableDataStructureEvalNeoDirectedSparseGraphTwoHundredWLFGCAvg}%
& \UseMacro{TableDataStructureEvalNeoDirectedSparseGraphTwoHundredWLFROAvg}%
& \UseMacro{TableDataStructureEvalNeoDirectedSparseGraphTwoHundredWLFROFGCAvg}%
& \UseMacro{TableDataStructureEvalMemDirectedSparseGraphTwoHundredWLAvg}%
& \UseMacro{TableDataStructureEvalMemDirectedSparseGraphTwoHundredWLFGCAvg}%
& \UseMacro{TableDataStructureEvalMemDirectedSparseGraphTwoHundredWLFROAvg}%
& \UseMacro{TableDataStructureEvalMemDirectedSparseGraphTwoHundredWLFROFGCAvg}%
\\%
\midrule%
\multicolumn{1}{l}{\UseMacro{TableLocNocAvg}}%
& \UseMacro{TableDataStructureEvalNeoWLAvg}%
& \UseMacro{TableDataStructureEvalNeoWLFROAvg}%
& \UseMacro{TableDataStructureEvalNeoWLFGCAvg}%
& \UseMacro{TableDataStructureEvalNeoWLFROFGCAvg}%
& \UseMacro{TableDataStructureEvalMemWLAvg}%
& \UseMacro{TableDataStructureEvalMemWLFROAvg}%
& \UseMacro{TableDataStructureEvalMemWLFGCAvg}%
& \UseMacro{TableDataStructureEvalMemWLFROFGCAvg}%
\\%
\multicolumn{1}{l}{\UseMacro{TableLocNocSum}}%
& \UseMacro{TableDataStructureEvalNeoWLTotal}%
& \UseMacro{TableDataStructureEvalNeoWLFROTotal}%
& \UseMacro{TableDataStructureEvalNeoWLFGCTotal}%
& \UseMacro{TableDataStructureEvalNeoWLFROFGCTotal}%
& \UseMacro{TableDataStructureEvalMemWLTotal}%
& \UseMacro{TableDataStructureEvalMemWLFROTotal}%
& \UseMacro{TableDataStructureEvalMemWLFGCTotal}%
& \UseMacro{TableDataStructureEvalMemWLFROFGCTotal}%
\\%
\bottomrule%
\end{tabular*}%
\vspace{2pt}%
\end{table*}%

Furthermore, we benchmarked the implemented methods for our two
prototypes with different optimizations for random workloads (number of elements
in the data structures were randomly varied between 10-500 for every run) and
the results are shown in Table~\ref{tab:table_data_structure}.
The reported execution times are averaged over 200
runs (We also compared the outcome of \Tool implementation
  with imperative implementation for each run for further checking
  correctness of \Tool). We see that \ToolMemory once again
outperforms \ToolNeo, furthermore, we also observe certain instances
where test execution times of \ToolNeo is significantly larger. Based
on the profiled data, this stems from the unpredictability in
execution times of node and relationship creations.

A detailed breakdown of the major steps involved in the \ToolNeo flow
is given in Figure~\ref{figure:step_breakdown}. We see that the optimizations
\textbf{+WL} and \textbf{+WL+FRO} are comparable in performance because, 
although \textbf{+FRO} limits the \emph{Follow references} step to reporting
only objects transitively reachable from the passed in root object, the entire
object graph is still traversed just not reported in the callback as per JVMTI
specifications. Steps such as \emph{Tag zero} and \emph{Assign unique tag} that
depend on the total number of objects on the heap are greatly impacted by forcing 
a garbage collection event through \textbf{+FGC}.

\section{Discussion and Future Work}
\label{sec:limitations}

Although \Tool is already production ready, there are endless
opportunities for improvements and applications.
We document several directions in this section.

\MyPara{Performance}
We have focused our work on implementing various features rather than
\Tool's performance, so far.  There is substantial work to be done to
get query performance with \Tool closer to equivalent imperative
implementations.
Future work should implement various optimizations that most graph
databases already include, such as graph compression
techniques~\cite{Fan_2012,Feder_1991},
indexing~\cite{Chen_2021,Yan_2005}, memory-efficient custom data
structures~\cite{Van_DS,Vitter_2001,Hellings_2012}.

\MyPara{Query language}
Currently, \Tool uses \Cypher as the query language. There are several other popular
alternatives that can be supported in the future, including 
Gremlin~\cite{Gremlin_QL}, SPARQL~\cite{SPARQL}, PGQL~\cite{PGQL}, 
GSQL~\cite{GSQL}, and GraphQL~\cite{GraphQL}.
We chose \Cypher as it is the most dominant graph query 
language at this point. Additionally, we were very much familiar with \Neo.

\MyPara{Debugging}
In this paper, we focused the evaluation solely on
having \Tool being used to program an application. Another direction
is to bring \Tool to support development tools. In this
  context, we have preliminarily used \Tool to query program state
  within a debugging session of \Code{jdb}~\cite{JDB}. Our objective
  was to identify object confinement of the edges and vertices
  (\Code{Integer}) of a \Code{SparseGraph} (\jung) instance.
We first created a \Code{SparseGraph} instance with one of its vertices 
being referenced by another object outside the confinement of the 
\Code{SparseGraph} instance. We next stopped the program execution at a 
breakpoint using \Code{jdb} and executed an \Tool query within the 
\Code{jdb} session using \Code{jdb}'s \Code{eval} command to check 
ownership of the vertex. The \Tool query was successful in identifying a 
reference chain to the vertex from the object outside the confinement of the
\Code{SparseGraph} instance.
Further integration with such
  debugging environments and IDEs would enable developers to
navigate the entire state of a program in an easy way (by writing
\unboundedqs) and discover interesting values and relations.
Finally, having data in a graph database
already provides data visualization capabilities with
off-the-shelf tools; existing graph visualization libraries are way
more advanced than any existing visual
debugger~\cite{Oechsle_VisualDebugger, Alsallakh_VisualDebugger,
  GuETAL14CapturingIDEInteractions, DDD}.

\MyPara{Snapshots}
Furthermore, \ToolNeo serializes object graphs as a part of
  its flow and hence, essentially captures a snapshot of the
  heap. This can be used to analyze differences of the
  heap. More excitingly, this enables (1)~time travel
  debugging~\cite{Expositor_Yit, Tardis_Earl,TTD_Earl}, a powerful
  debugging technique that allows tracking of the sequence of program
  states leading to the error, and (2)~identifying memory leaks
  ~\cite{MemoryLeak_Evan, MemoryLeak_Markus,
    MemoryLeak_AntTracks_Markus}.

\MyPara{Safety}
\Tool allows developers to break one of the core software engineering
principles: encapsulation.  While the power of \Tool enables various
ways to treat the system, responsible use has a great potential.
Furthermore, there are ways in which encapsulation in Java (and other
languages) is already being broken (e.g.,
\Code{Unsafe}~\cite{Huang_Unsafe, Mastrangelo_Unsafe}) when it comes
to designing program analysis tools.  Having another, more effective
way to implement analyses tools, is a plus.

\MyPara{Languages}
\Tool idea is applicable beyond Java and integration with other
languages, especially those that are dynamically typed, is a planned
future work.

\section{Related Work}

We cover the most closely related work in this section by comparing
our work with the following groups: (1)~program analysis using query
and domain specific languages (DSL), and (2)~minimizing impedance
mismatch between imperative programming languages and database
systems.

\MyPara{Program analysis} Closely related to \Tool are \emph{Fox}~\cite{foxObjectGraph} and \emph{Datalog}~\cite{Ceri_datalog}.
Fox uses a DSL to analyze object graph in the \JVM heap for
aliasing, confinement and ownership. Datalog and its applications to (static)
program analysis have been explored in numerous studies~\cite{Huang_datalog,Kolaitis_datalog, Cali_datalog}.
Most prominently, the Doop framework~\cite{Smaragdakis_Doop} and followup
work~\cite{Smaragdakis_datalog} express various forms of pointer analysis
in Datalog by exploiting its expressive power.
We focus on dynamic program analysis and our key insight that the entire
program heap can be seen as a single graph database which can be
queried via popular graph query languages.
The expressive power of \Cypher as a query language enables concise
descriptions in applications as shown in Section~\ref{sec:example}.

\MyPara{Impedance mismatch}
\emph{Impedance Mismatch}~\cite{Maier_ImpMismatch} refers to the 
friction of interfacing imperative languages with database systems.
Efforts to identify, categorize~\cite{Cook_IntegratingPL,Keller_PL_DB_Unification,Copeland_SmallTalk_ImpMismatch} and reduce this mismatch have been
achieved through object-oriented databases, object-relational mappers,
data access APIs, embedded query languages~\cite{Cook_IntegratingPL} and language
integrated queries.

\vspace{3pt}
\MyParaTwo{Call level interface (CLI)} are API's such as
JDBC~\cite{HamiltonJDBC} and ADO.Net~\cite{Adya_ADO,
  Castro_ADO,Blakeley_ADO, Chaudhuri_ADO_Performance} that abstract
away the generation of the query language through API
methods. However, it is difficult~\cite{LawrenceJDBCPerformance} to
ensure the efficiency of the generated queries. \emph{JCypher}
~\cite{JCypher} is an example of a CLI for \Cypher.

\vspace{3pt}
\MyParaTwo{Embedded query languages} API's such as
SQLJ~\cite{Eisenberg_SQLJ_1, Eisenberg_SQLJ_2} and XJ~\cite{Harren_XJ,
  Harren_XJ_Performance} allow embedding the query language to query
external databases. Although \Tool shares similarities in that it
allows writing queries in \Cypher inside \Java, however, \Tool allows
the in-memory object graph to be queried.

\vspace{3pt}
\MyParaTwo{Object oriented databases} Object oriented databases (OOD)
~\cite{Banciihon_OODBMS, Carey_OODBMS,Kopteff_ORM_VS_OODBMS, Straube_OODBMS, Maier_OODBMS_OPAL}
have been introduced to couple object-oriented programming languages and
databases.
OOD is more about persistence of objects rather than being able to
query relations and object graph available at runtime, which \Tool enables.

\vspace{3pt}
\MyParaTwo{Relational object mappers}
Relational object mappers~\cite{Linskep_ORM, Burza_ORM, Torres_ORM,
  Cheung_ORM, Neil_ORM_Hibernate}, like Hibernate, enable conversion
of data between type systems.  They convert objects to (relational)
database by automatically grouping properties and enable loading and
updating these values.
\Tool is about querying object graphs not about persistence. The
closest connection with object mappers is our translation from an
object graph into a graph database.

\vspace{3pt}
\MyParaTwo{Language integrated queries}
(\LINQ)~\cite{MeijerETAL06LINQ}, developed by Microsoft, is a
technology that adds native data querying capabilities to .NET
languages.  As a data source, \LINQ can use in-memory data, i.e., any
collection that implements \Code{IEnumerable} (e.g., \Code{List},
\Code{SortedSet}).  Although powerful, \LINQ provides no
support to query arbitrary objects and their relations. Language
integrated queries have seen renewed interest~\cite{OhoriETAL11SML,
  CheneyETAL13LINQTheory, NagelETAL14Codegen, SuzukiETAL16SafeLINQ,
  KowalskiETAL17QueryUnnesting, KiselyovETAL17SoundLINQ,
  SalvaneschiETAL19DistributedQueries, LopezETAL20Optica,
  OkuraETAL20OptimizingQueries, RicciottiETAL21QueryLifting} since the
release of the LINQ framework. Similar frameworks for \Java include
SBQL4J~\cite{sbql4j} and Quaere~\cite{quaere}.

\section{Conclusion}

We introduced \ogo (\Tool), a novel paradigm that combines imperative
(object-oriented) programming and declarative queries.  \Tool treats
the program state (i.e., object graph) as a graph database that can be
queried and modified using graph query language(s); \Tool currently
uses \Cypher as the primary query language.
Each object in an object graph is a node, each primitive, String and
primitive array field is a property, and each reference field forms a
relation between two nodes.
\Tool is ideal for querying collections (similar to \LINQ),
introspecting the runtime system state (e.g., finding all instances of
a given class or accessing fields via reflection), and writing
assertions that have access to the entire program state.
We prototyped \Tool for Java in two ways: (a)~by translating the \JVM
heap object graph into a Neo4j database on which we run \Cypher
queries, and (b)~by implementing our own in-memory graph query engine
that directly queries the object graph.
We used \Tool to rewrite hundreds of statements in large open-source
projects into \Tool queries.
Our evaluation shows the wide applicability of our approach and good
first results with in-memory implementation.
\Tool enables an entirely different view of objects and data, which
will move programming experience to the next level.

\section*{Acknowledgements}

We thank Rifat Seleoglu for his contribution in writing assertions
using \Tool and Nader Al Alwar, Joseph Kenis, Pengyu Nie, Yu Liu,
Zhiqiang Zang, Jiyang Zhang, and the anonymous reviewers for their
comments and feedback.
This work is partially supported by the US National Science Foundation
under Grant CCF-2107206, CCF-2107291, and CCF-2217696.


\onecolumn
\begin{multicols}{2}
\bibliographystyle{ACM-Reference-Format}
\interlinepenalty=10000
\bibliography{bib}
\end{multicols}

\end{document}